\documentclass[12pt]{iopart}
\usepackage{cite}
\usepackage{graphicx}
\def\be{\begin{equation}}
\def\ee{\end{equation}}
\def\omegac{\omega_{\rm c}}
\def\omegazeroone{\omega_{01}}
\def\Ec{E_{\rm c}}
\def\EJ{E_{\rm J}}
\def\ng{n_{\rm g}}

\begin{document}
\title[Circuit QED]{Circuit QED and engineering charge based superconducting qubits}
\author{SM Girvin, MH Devoret, and RJ Schoelkopf}
\address{Department of Applied Physics, Yale University, P.O.~Box 208284
, New Haven, CT 06520-8248}
\eads{\mailto{steven.girvin@yale.edu}, \mailto{michel.devoret@yale.edu}, \mailto{robert.schoelkopf@yale.edu}}
\begin{abstract}
The last two decades have seen tremendous advances in our ability to generate and manipulate quantum coherence in mesoscopic superconducting circuits.  These advances have opened up the study of quantum optics of microwave photons in superconducting circuits as well as providing important hardware for the manipulation of quantum information.  Focusing primarily on charge-based qubits, we provide a brief overview of these developments and discuss the present state of the art.  We also survey the remarkable progress that has been made in realizing circuit QED in which superconducting artificial atoms are strongly coupled to individual microwave photons.
\end{abstract}
\submitto{Phys. Scr. T 137 (2009) 014012}
\maketitle
\section{Introduction}
The remarkable recent progress in creating superconducting quantum bits and manipulating their states has been summarized in several reviews \cite{Devoret2004,Esteve2005,Wendin_Shumeiko2006,Wendin_Shumeiko2007,Clarke2008,WiringUpQuantumSystems,YouandNori2005,Nori2008,Korotkov2009}.  Nearly 30 years ago Leggett discussed the fundamental issues concerning the collective degrees of freedom in superconducting electrical circuits and the fact that they themselves can behave quantum mechanically \cite{Leggett1980}.  The essential collective variable in a Josephson junction \cite{MartinisDevoretJJreview} is the phase difference of the superconducting order parameter across the junction.  The first experimental observation of the quantization of the energy levels of the phase `particle' was made by Martinis, Devoret and Clarke in 1985 \cite{MartinisDevoretClarke1985, Clarke1988Science}.

Caldeira and Leggett also pointed out the crucial role that quantum fluctuations of the dissipative electromagnetic environment play in the quantum coherence and dynamics of Josephson junctions \cite{CaldeiraandLeggett1983}. These ideas were tested in early experiments on macroscopic quantum tunneling (MQT) \cite{VossandWebb1981,DevoretMartinisClarkeMQT1985}.  In a novel experiment, Turlot \etal \cite{Turlot1989} used a sliding absorber to mechanically vary the electromagnetic impedance seen by the Josephson junction and hence modulate the rate of MQT.  It was subsequently realized that the quantum fluctuations of the environment can play an important role even in transport through normal metal junctions \cite{DevoretPofE,GirvinPofE}.

This physics of the environment is related to that of the Purcell effect \cite{Purcell1946}, first observed for a superconducting qubit by Houck \etal \cite{HouckPurcell2008}.  Here a qubit placed inside a cavity can have its decay rate suppressed if it is far detuned from the cavity resonance or enhanced if the qubit transition frequency is close to the cavity resonance.  The former is useful for protecting quantum superpositions.  The latter is useful for providing rapid qubit reset to the ground state.  It has also been successfully used to generate single microwave photons on demand and enhance the fidelity of coherent quantum information transfer from a superconducting qubit to a `flying' photon qubit \cite{Houck2007SinglePhoton}.  One can view the Purcell effect as the resonator performing an impedance transformation on the external dissipation presented by the environment to the qubit \cite{HouckPurcell2008}. Neeley \etal \cite{Neeley2008a} have used a tunable transformer coupling to quantitatively explore the role of environmental coupling in a phase qubit circuit over a wide range of coupling strengths.

\section{Circuit QED}

Quantum electrodynamics (QED) is the study of the interaction of atoms with the quantized electromagnetic field.  In cavity QED one modifies the electromagnetic environment by placing the atoms inside a high finesse Fabry-P\'erot resonator.  This not only simplifies the physics by making the spectrum of electromagnetic modes discrete, it also gives the experimentalist control over both the damping of the resonances and their detuning from the atomic transition frequency.  Furthermore, because the photons bounce between the mirrors many times, their coupling to the atoms can be greatly enhanced.  In the simplest approximation (including the rotating wave approximation), the system is described by the Jaynes-Cummings Hamiltonian
\be
H= \hbar\omegac a^\dagger a + \frac{\hbar\omegazeroone}{2}\sigma^z + \hbar g\left\{a\sigma^+ + a^\dagger\sigma^-\right\} + H_{\rm drive} + H_{\rm damping},
\ee
where the single cavity mode is described as a simple harmonic oscillator of angular frequency $\omega_{\rm c}$, the two-level atom is represented as a simple spin-1/2 with excitation frequency $\omega_{01}$, and the `vacuum Rabi coupling', $g$, represents the dipole matrix element for the process in which the atom absorbs or emits a photon.  The external driving and damping terms, not written explicitly here, which help control the electromagnetic state of the cavity, are treated using the input-output formalism of quantum optics \cite{Clerk2008}.  The extension of this Hamiltonian to the case of multiple qubits is known as the Tavis-Cummings model \cite{Tavis1968}.

There is a long history of cavity QED studies in the AMO community for both alkali atoms in optical cavities \cite{Mabuchi02,Walls94,Thompson1992,Boca2004,SchusterAMOVacRabi2008} and Rydberg atoms in microwave cavities \cite{Nogues1999,Guerlin2007,Gleyzes2007,Deleglise2008,HarocheRaimondRMP,HarocheRaimondcQEDBook}.  In the optical case one typically monitors the effect of the atoms on the photons transmitted through the cavity.  It is not possible to measure the state of the atoms after they have fallen through the cavity because the spontaneous emission lifetime is on the order of nanoseconds at optical frequencies.  In the microwave experiments pioneered by the Paris group it is difficult to directly measure the microwave photons but relatively easy to measure the state of the Rydberg atoms with very high fidelity after they exit the cavity since they have a lifetime of approximately 30 ms and can be probed with state-selective ionization.

\begin{figure}[htp]
\centering
\includegraphics[clip]{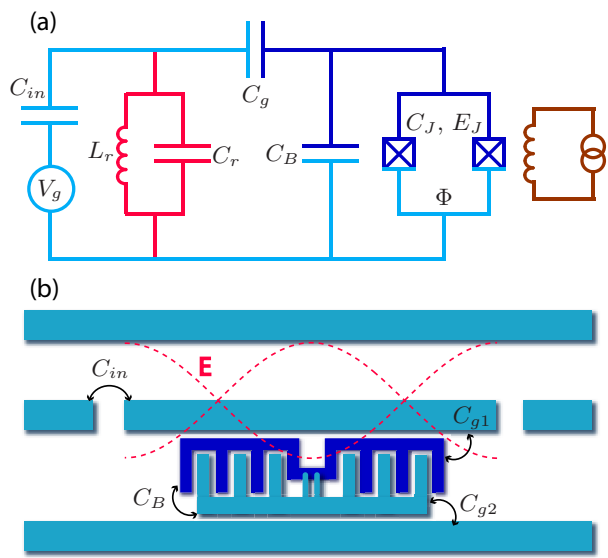}
\caption{Circuit QED: Schematic illustration (not to scale) of a transmon qubit embedded in a coplanar waveguide resonator. Panel (a) shows the lumped element circuit equivalent to the distributed circuit shown in panel (b).
From \cite{Koch2007}.}\label{fig:transmonschematic}
\end{figure}

`Circuit QED' uses superconducting qubits as artificial atoms coupled to microwave resonators \cite{BlaisCQEDtheory2004,WallraffCQED2004,Chiorescu2004,Devoret2007,WiringUpQuantumSystems} as illustrated schematically in figure \ref{fig:transmonschematic}.  Measuring the amplitude and phase of microwaves transmitted through the resonator realizes the equivalent of optical cavity QED at microwave frequencies.  In recent years there were many theoretical proposals for coupling qubits to either three-dimensional cavities or lumped element resonators \cite{Shnirman97,Makhlin01,Buisson01,Marquardt01,Al-Saidi01,Plastina03,Blais03,Yang03,you03a} and there has been a flurry of experiments
\cite{WallraffCQED2004,Chiorescu2004,Schuster2005,Wallraff2005,Schuster2007a,
Johansson2006,Siddiqi2006,Boulant2007,
Schuster2007,Wallraff2007,Sillanpaa2007,Houck2007SinglePhoton,
Leek2007,Majer2007,Astafiev2007,Metcalfe2007,Deppe2008,Fink2008,Hofheinz2008,Wang2008,
Schreier2008,HouckPurcell2008,Fragner2008,Grajcar2008,Sandberg2008,
Il'ichev2009,Bishop2009,Chow2009,Hofheinz2009,DiCarlo2009}
and further theoretical proposals too numerous to list.  The beauty of coplanar waveguide resonators is that they are quasi-one-dimensional Fabry-P\'erot cavities with orders of magnitude smaller mode volume than can be achieved with ordinary three-dimensional resonators.  Because the mode volume of these quasi-one-dimensional resonators can be as small as $10^{-6}$ cubic wavelengths, and because the artificial atoms have transition dipoles much larger than even Rydberg atoms, the coupling strength $g$ between the atom and a single photon in the resonator is enormously enhanced and becomes orders of magnitude larger than can be ordinarily achieved.  In fact the dimensionless ratio of the coupling strength to the transition frequency approaches the limit set by the fine structure constant $\alpha$ \cite{WiringUpQuantumSystems}
\be
\frac{g}{\omegazeroone} \sim \sqrt{\frac{\alpha}{\epsilon}},
\ee
where $\epsilon$ is the dielectric constant of the medium surrounding the qubit.  (Strictly speaking this limit is obtained assuming that the quantum charge fluctuations are only of order one Cooper pair.  In actuality they increase slowly as $(\EJ/\Ec)^{1/4}$ so this limit can be exceeded.) For the lumped element equivalent circuit shown in figure \ref{fig:transmonschematic} the vacuum Rabi coupling is given by
\be
g=\frac{C_{\rm g}}{2}\sqrt{\frac{\omega_{\rm q}\omega_{\rm r}}{C_{\rm q}C_{\rm r}}},
\ee
where $\omega_{\rm q}$ is the qubit frequency, $\omega_{\rm r}$ is the resonator frequency, and $C_{\rm q}=C_{\rm B}+2C_{\rm J}$ is the capacitance across the the Josephson junction.  The values of the lumped circuit elements are determined by the capacitance matrix of the actual distributed circuit elements \cite{Koch2007}.

There exists a dual geometry in which the Josephson junction qubit is placed in line with the center pin of the resonator and couples directly to the microwave currents flowing in the resonator \cite{Devoret2007,Fluxqubitstrongcoupling2009}.  In this dual geometry the fine structure constant is replaced by its \emph{inverse} and the problem is engineering the circuit to reduce the coupling to manageable levels.

The coupling $g$ is most readily measured by tuning the qubit transition frequency $\omega_{01}$ to match the cavity frequency $\omega_{\rm c}$.  The resulting degeneracy is lifted by the dipole coupling term leading to the so-called vacuum Rabi splitting $2g$ \cite{Thompson1992,HarocheRaimondRMP,WallraffCQED2004,Boca2004,Johansson2006,SchusterAMOVacRabi2008}. The two lowest lying excited states are coherent superpositions (`bonding-anti-bonding' combinations) of photon excitation and qubit excitation.  The coupling available in circuit QED is now so strong that splittings of $\sim 300$ line widths are easily achieved \cite{Fink2008,Bishop2009}.  The higher lying excited states form a strongly anharmonic ladder which can be explored by either strong driving or use of two excitation tones \cite{Deppe2008,Fink2008,Bishop2009}.

In the so-called dispersive regime where the qubit is far detuned from the cavity ($|\omegazeroone-\omegac|\gg g$), diagonalization of the Hamiltonian to lowest order in $g$ leads to a second-order dispersive coupling which is QND with respect to both the photon number and the qubit energy
\be
V=\hbar\frac{g^2}{\Delta}[a^\dagger a+\frac{1}{2}] \sigma^z,
\label{eq:dispersivecoupling}
\ee
where $\Delta\equiv\omegazeroone-\omegac$ is the detuning of the qubit from the cavity.  The dispersive coupling can be interpreted either as a shift in the cavity frequency which depends on the state of the qubit, or as the `ac-Stark' or `light' shift (plus the Lamb shift \cite{BlaisCQEDtheory2004,Schuster2005,Fragner2008}) of the qubit frequency proportional to the number of photons in the cavity.  The qubit-state-dependent shift of the cavity frequency leads to changes in the amplitude and phase of photons reflected from or transmitted through the cavity and is the basis of the QND readout of the qubit state in circuit QED \cite{BlaisCQEDtheory2004,WallraffCQED2004}.   The mean value of the light shift can be used to rapidly tune qubit transition frequencies \cite{Schuster2005,Schuster2007a,Gambetta2006,Majer2007}.  The fluctuating part of the light shift can be viewed as the quantum back action \cite{Clerk2008} of the qubit measurement.  As required by the principles of quantum measurement \cite{Clerk2008}, the photon shot noise \cite{Bertet2005} in the cavity gradually dephases the qubit superposition as information is gained about $\sigma^z$.  This back action effect leads to a broadening of the spectroscopic line width of the qubit \cite{Schuster2005,Schuster2007a,Ithier2005,Lupascu2005,Boissonneault2008,Boissonneault2009}.  In the so-called `strong-dispersive' regime \cite{Schuster2007}, the coupling is so large that the light shift per photon exceeds both the cavity line width $\kappa$ and the atom line width $\gamma$: $\frac{g^2}{\Delta}> \kappa,\gamma$.  In this regime the qubit spectrum breaks up into a series of separately resolved peaks representing the distribution of photon numbers within the driven cavity \cite{Schuster2007}.  This `photon number' detector was used to distinguish thermal and coherent states in the cavity and could be used to measure number-squeezed states and other non-classical states \cite{Schuster2007}.  This strong-coupling physics has been beautifully observed in the time domain by the Paris group \cite{Nogues1999,Guerlin2007,Gleyzes2007,Deleglise2008}.

The rate of progress in observing novel strong coupling non-linear quantum optics effects in superconducting electrical circuits is quite remarkable.  As noted above, Houck \etal used the Purcell effect \cite{HouckPurcell2008} to generate non-classical photon states in a cavity \cite{Houck2007SinglePhoton}.  The states were a superposition of $n=0$ and $n=1$ Fock states with controlled amplitude and phase.  `Fluorescence tomography' was performed on these states using square law detection to determine the probability of having a photon.  In addition, homodyne measurements were performed to determine the two quadratures of the electric field which are controlled by the off-diagonal coherence between the $n=0$ and $n=1$ Fock states.  In particular they showed that the mean electric field of the one-photon Fock state was zero.

Higher Fock states up to $n=6$ were synthesized by the UCSB group \cite{Hofheinz2008} who also observed that the decay rate scaled linearly with $n$ as expected \cite{Wang2008}.  This same effect was seen qualitatively in the frequency domain in the experiment of Schuster \etal \cite{Schuster2007}. The qubit spectrum showed up to 6 resolved peaks displaying the distribution of photon numbers within the driven cavity and the line width of the peaks increased with $n$.  In a 2009 tour-de-force, Hofheinz \etal \cite{Hofheinz2009} demonstrated a remarkable method for synthesizing arbitrary photon states (including Fock and various cat states) in a cavity and measuring their Wigner distributions.  This level of control now exceeds what has been possible to date with atomic physics methods.

Because microwave photons have $10^4$ to $10^5$ times less energy than visible photons, they are much more difficult to detect.  The work of Houck \etal \cite{Houck2007SinglePhoton} and Schuster \etal \cite{Schuster2007} showed that individual photons could be detected with low efficiency and the recent work of Hofheinz \etal \cite{Hofheinz2009} demonstrated very high efficiency detection of individual photons in a cavity. However a general purpose high bandwidth `photomultiplier' does not yet exist in the microwave regime.  There have been some theoretical proposals for single photon detection \cite{Helmer2009a,Romero2009} but this remains an important open experimental problem.

Another novel new direction is construction of single artificial atom `lasers' \cite{Astafiev2007,Marthaler2008,Ashhab2009} as well as Sisyphus cooling and amplification \cite{Grajcar2008} of an oscillator.  The extreme strong coupling available should permit observation of `photon blockade' effects \cite{Birnbaum2005}, and parametric down-conversion by three-wave mixing \cite{Moon2005,Marquardt2007}.  The advances in our understanding and fabrication of Josephson junction circuits motivated by the quest for a quantum computer have led to dramatic advances in the ability to do four-wave mixing, parametric amplification near the quantum limit, as well as strong squeezing of the vacuum \cite{Castellanos-Beltran2008,Bergeal2008}.  These advances will not only permit much better dispersive readout of qubits, they also open up the possibility of continuous variable quantum information processing \cite{Gottesman2001,Braunstein2005} since two-mode squeezed states are an entanglement resource.

\section{Charge-Based Qubits and Variations Thereof}

Artificial atoms can be constructed from electrical circuit elements \cite{Esteve2005,MartinisDevoretJJreview}.  Clearly we want to avoid explicit resistors, which can be in the form of dielectric losses in the tunnel junctions and substrate \cite{Martinis2005} as well as the `radiation resistance' represented by coupling to transmission lines.  The simplest (ideally) purely reactive circuit elements are inductors and capacitors, but these can only be used to construct a harmonic oscillator whose evenly spaced energy levels are not suitable for making qubits.  We must incorporate a non-linear circuit element.  The only known non-linear circuit element which is also non-dissipative is the Josephson junction \cite{MartinisDevoretJJreview}.  A number of different qubit designs have been developed around the Josephson junction including the Cooper pair box \cite{Averin1985,Buttiker1987,Lafarge1993,BouchiatCPBPhysScr1998,Nakamura1999,Vion2002,Koch2007,Schreier2008} based on charge, the flux qubit \cite{Mooij1999,Wal2000,Chiorescu2003}, and the phase qubit \cite{Martinis2002,Berkley2003}.  The Cooper pair box is topologically distinct from the other two designs in that it has no wire closing the loop around the junction.  Hence the number of Cooper pairs transferred through the junction is a well-defined integer.  The integer charge implies the conjugate phase is compact; that is, in the phase representation, the system obeys periodic boundary conditions.  As we will see below, this implies that charge-based qubits are sensitive to stray electric field noise, but this can be overcome.

\subsection{The Cooper Pair Box}

\begin{figure}[htp]
\centering
\includegraphics[clip]{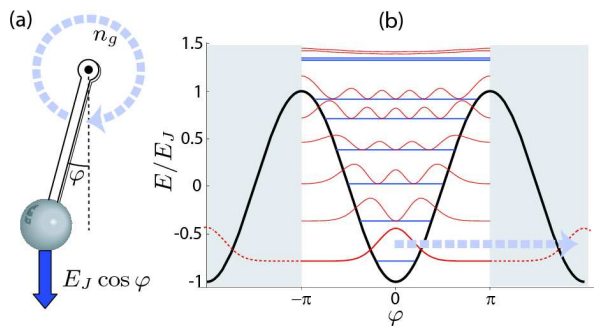}
\caption{The Cooper pair box Hamiltonian in the phase representation is equivalent to that of a quantum rotor.  The offset charge $\ng$ is equivalent to an Aharonov-Bohm flux which produces a Berry phase proportional to the winding number of the rotor trajectory.  Unlike other qubit circuit topologies, the rotor wave function obeys periodic boundary conditions.  From \cite{Koch2007}.}
\label{fig:quantumrotor}
\end{figure}

\begin{figure}[htp]
\centering
\includegraphics[clip]{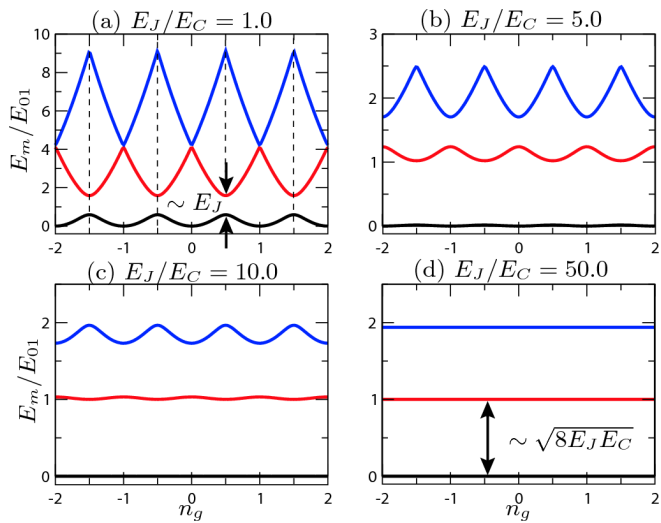}
\caption{Energy spectrum of the Cooper pair box as a function of offset charge for different values of the dimensionless ratio of Josephson energy to charging energy.  The exponential decrease in the charge dispersion is clearly seen. From \cite{Koch2007}.}
\label{fig:CPBspectrum}
\end{figure}

The Cooper Pair Box (CPB) Hamiltonian is given by
\be
H= 4\Ec\left[\hat n -\ng\right]^2 -\EJ\cos{\hat \varphi}
\ee
where $\hat n$ is the integer-valued Cooper pair number operator and $\ng$ is a continuous valued offset charge  (or `gate charge') representing dc bias intentionally applied to the qubit, low frequency stray electric fields in the system (`charge noise') as well as high frequency electric fields from photons in the cavity in which the qubit is placed.  $4\Ec$ is the charging energy for a Cooper pair and $\EJ$ is the Josephson tunneling energy.
In the phase representation $\hat n \longrightarrow -i\frac{\partial}{\partial \varphi}$ and the wave function obeys $\Psi(\varphi+2\pi)=\Psi(\varphi)$.  As illustrated in figure \ref{fig:quantumrotor}, this corresponds to the Hamiltonian of a quantum rotor with moment of inertia controlled by the charging energy and gravitational potential controlled by the Josephson energy.  $\hat n$ plays the role of the integer valued angular momentum of the rotor and the torque associated with the cosine potential changes the angular momentum up and down by one unit.  The offset charge $\ng$ appears as a vector potential that induces an Aharnov-Bohm phase proportional to the winding number of the rotor's trajectory.
Numerical diagonalization is more readily done in the charge basis where the Josephson term is tri-diagonal:  $\langle n\pm1|\cos \varphi|n\rangle= \frac{1}{2}$.  Clearly the qubit spectrum is periodic in $\ng$ with unit period as can be seen in figure \ref{fig:CPBspectrum}.

The first evidence that Josephson tunneling causes the Cooper pair box to exhibit coherent superpositions of different charge states was obtained by Bouchiat \etal \cite{BouchiatCPBPhysScr1998}.  This was followed in 1999 by the pioneering experiment of the NEC group \cite{Nakamura1999} demonstrating time-domain control of the quantum state of the CPB using very rapid control pulses to modulate the offset charge.

For generic values of the gate charge, the ground state $\Psi_0$ and excited state $\Psi_1$ differ in their respective static electric `dipole moments'
\be
p_j\sim 2ed \langle \Psi_j|\hat n|\Psi_j\rangle
\ee
where $d$ is (approximately) the distance between the two islands of the qubit.  (More precisely, the effective value of $d$ depends in a complex way on the details of the cavity and qubit geometry and the resulting capacitance matrix \cite{Koch2007}.) Nakamura \etal \cite{Nakamura1999} used the dependence of a certain quasi-particle tunneling rate on $p_j$ to readout the state of the qubit.  Aassime \etal \cite{Aassime2001} and Lehnert \etal  \cite{Lehnert2003} developed an RF single electron transistor readout scheme for charge based qubits.

Unfortunately in the regime where charge based readout works, a stray electric field ${\mathcal E}$ causes a first-order perturbation theory shift of the qubit excitation frequency by an amount
\be
\delta \omegazeroone=\frac{1}{\hbar}{\mathcal E}(p_1-p_0).
\ee
This leads to very rapid dephasing of quantum superpositions at rate \cite{Martinis2003}
\be
\frac{1}{T_\varphi}=\frac{1}{2}\left(\frac{p_1-p_0}{\hbar}\right)^2S_\mathcal{EE},
\ee
where $S_\mathcal{EE}$ is the electric field spectral density at low frequencies.  The total decoherence rate is then given by
\be
\frac{1}{T_2^*} = \frac{1}{2T_1}+\frac{1}{T_\varphi}.
\ee

\subsection{The Quantronium Qubit}

The next great advance was the first Ramsey fringe experiment in an electrical circuit performed at Saclay \cite{Vion2002} using a charge qubit dubbed the quantronium.  This group recognized that there is a sweet spot in offset charge at $\ng=1/2$ for which the ground and excited state have no difference in dipole moment:  $p_1=p_0$. At this bias point, the energy splitting is an extremum with respect to offset charge.  Hence the transition frequency depends on stray electric fields only in second order
\be
\delta\omegazeroone = \frac{\partial \omegazeroone}{\partial\ng}\bigg|_{\ng=1/2} \left(\ng-\frac{1}{2}\right)
+\frac{1}{2}\frac{\partial^2 \omegazeroone}{\partial\ng^2}\bigg|_{\ng=1/2} \left(\ng-\frac{1}{2}\right)^2+...
\label{eq:domega}
\ee

By analogy with the effects of stray magnetic fields on atomic transitions and the optimal working point used in atomic clocks, the charge noise sweet spot is sometimes called the `clock point'.  Working at this point, the Saclay group  managed to extend $T_2^*$ approximately three orders of magnitude to $\sim 500$ns.  Subsequently at Yale, the Devoret group \cite{Metcalfe2007} was able to perform very rapid sequences of Ramsey fringe experiments and show that the measured decoherence rate was consistent with the second-order effects of the curvature of the transition frequency in the vicinity of the extremum, given typical values for the offset charge noise.

We now arrive at an interesting quandary.  By tuning the qubit to the sweet spot, the environment is no longer able to detect which state the qubit is in, based on coupling to its electric dipole moment.  This is why the coherence time is so dramatically enhanced.  But, if the environment cannot measure the state of the qubit by looking at the dipole moment, neither can we!  The Saclay group recognized this and developed the concept of reading out the qubit by measuring the state dependent susceptibility (inductance).  Rather than going into the details of this, it is easier in the context of the present discussion to instead explain the closely related state-dependent susceptibility method based on capacitance developed by the Yale group \cite{BlaisCQEDtheory2004,WallraffCQED2004}.  Because the offset charge is essentially equivalent to an applied voltage, and the potential energy of a capacitor is $\frac{1}{2}CV^2$, the second derivative of the transition energy with respect to $\ng$ in (\ref{eq:domega}) is essentially the difference in quantum capacitance \cite{Widom1984,Averin1985,LiharevandZorin1985,Averin2003,Duty2005,Sillanpaa2005} presented to an external probing field when the qubit is in the ground and excited states.  Essentially this effect was used by the Yale group in developing the dispersive readout \cite{BlaisCQEDtheory2004,WallraffCQED2004} based on (\ref{eq:dispersivecoupling}).  Working with a low frequency probe, precisely this effect was measured in \cite{Duty2005,Sillanpaa2005}.  The difference is that the high-frequency dispersive probe depends on the matrix elements related to the quantum capacitance, but as is clear from (\ref{eq:dispersivecoupling}), it also depends on the detuning of the qubit and resonator frequencies.  The importance of this difference will become clear below.

\subsection{Transmon Qubits}

\begin{figure}[htp]
\centering
\includegraphics[clip]{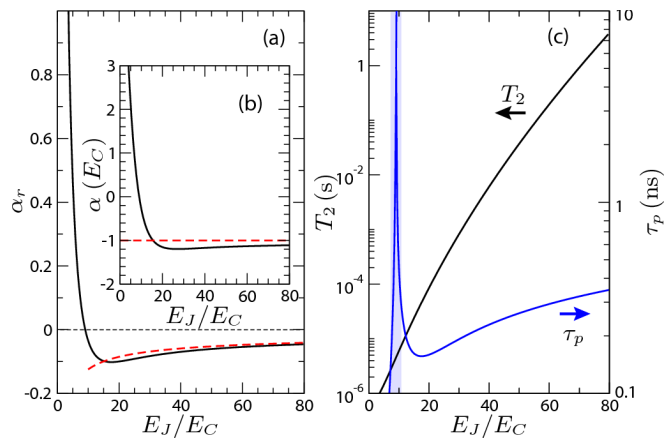}
\caption{a) Relative anharmonicity, $\alpha_{\rm r}$, of the CPB as a function of the dimensionless ratio of the Josephson and charging energies; b) absolute anharmonicity, $\alpha$, which asymptotically approaches the negative of the charging energy $\Ec$; c) (left axis) Phase coherence time $T_2$ given by the inverse of the charge dispersion, and (right axis) the minimum control pulse duration $\tau_{\rm p}$ set by the absolute anharmonicity.  The anharmonicity passes through zero at $\EJ/Ec=9$ and in this vicinity the CPB cannot be used as a two-level qubit since $\tau_{\rm p}$ diverges. From \cite{Koch2007}.}
\label{fig:anharmonicity}
\end{figure}

The most recent evolution of the charge qubit has been the `transmon' qubit developed by the Yale group \cite{Houck2007SinglePhoton,Koch2007,Schreier2008,LifeAfterChargeNoise2009} and schematically illustrated in figure \ref{fig:transmonschematic}.  This is nothing more than a CPB operated in the regime of large $\EJ/\Ec$.  Figure \ref{fig:anharmonicity} shows both the anharmonicity of the CPB spectrum, defined as the difference of the first two transition frequencies $\alpha \equiv \omega_{12}-\omega_{01}$, and the relative anharmonicity, defined as $\alpha_{\rm r} \equiv \alpha/\omega_{01}$.  The anharmonicity is important because it sets the limit $\tau_{\rm p}$ on how short qubit control pulses can be to avoid mixing in higher energy levels outside the logical subspace \cite{Steffen2003,Wallraff2005,Lucero2008,Chow2009}.   For $\EJ/\Ec<9$ the CPB spectrum has anharmonicity  which is positive.  Near $\EJ/\Ec=9$ the first two transition frequencies are too close together for the CPB to be used as a qubit.  Beyond $\EJ/\Ec=9$ the CPB has negative anharmonicity whose asymptotic value approaches the charging energy $\alpha\rightarrow -\Ec/\hbar$.  Because the transition frequency approaches the Josephson plasma frequency $\Omega_{\rm pl}=\sqrt{8\EJ\Ec}$, the \emph{relative} anharmonicity $\alpha_{\rm r}\sim -\sqrt{\frac{\Ec}{8\EJ}}$ asymptotically approaches zero as the system comes closer and closer to being a simple harmonic oscillator.  Nevertheless the charging energy is easily controlled using the experimental geometry and can be conveniently set in the range of a few hundred MHz, large enough so that all but the very shortest control pulses do not excite the system out of the subspace of the two lowest states.  Remarkably, the `charge dispersion' $\epsilon$ (overall amplitude of the periodic variation in transition frequency with offset charge) becomes exponentially small at large $\EJ/\Ec$ \cite{Averin1985,Koch2007,Schreier2008,LifeAfterChargeNoise2009}
\be
\epsilon \sim e^{-\sqrt{8\EJ/\Ec}},
\ee
so that for $\EJ/\Ec$ greater than roughly 50, dephasing due to charge noise is negligible.  Instead of a single charge sweet spot, the levels are essential flat independent of offset charge and every spot is sweet.  Coherence times without spin echo as large as $T_2^*\approx 3\mu$s with $T_\varphi \ge 35\mu$s have been observed for the transmon \cite{LifeAfterChargeNoise2009}.

We now arrive at our next major quandary.  If the energy eigenvalues are essentially independent of the offset charge then neither we nor the environment can read the state of the qubit using either the dipole moment or the susceptibility (quantum capacitance) since neither is dependent on the quantum state.  While this explains the even longer coherence times of the transmon, we are left to wonder how it is that the dispersive readout still works even though quantum capacitance is zero in both states.  Recall that if the qubit were actually a perfect harmonic oscillator, the transition frequencies would not respond at all to changes in offset charge (displacement of the origin of the oscillator).  It is obvious from classical considerations that the susceptibility would be a constant (given by the inverse of the spring constant) independent of the state.  The oscillator is highly polarizable and responds strongly to slow variations in offset charge by being displaced, but this displacement to a new equilibrium position has no effect on the excitation spectrum.  As noted above, the transmon comes exponentially close to this ideal behavior and yet, the dispersive readout still works.  This is because, while the charge dispersion falls off exponentially, the transmon retains its anharmonicity.  As can be seen from (\ref{eq:dispersivecoupling}), the cavity pull due to the virtual polarization of the qubit is strongly dependent on the detuning between the qubit and cavity.  For the case of the multi-level transmon, the expression for the cavity pull has to be rederived, but the essential point is that the detuning for the $0\rightarrow 1$ transition is not the same as that for the $1\rightarrow 2$ transition and so the cavity pull is state dependent, provided that the cavity frequency is reasonably close to the qubit.  For a very low frequency cavity, we are back in the regime measuring the quantum capacitance where the effect is small.

We now face one final quandary associated with the relatively weak ($\alpha_{\rm r} \sim 5-10\%$) anharmonicity of the transmon.  In the limit of large detuning from the cavity ($\Delta>\Ec>T_1$), the difference in cavity pulls between the qubit excited state and ground state is well approximated by \cite{Koch2007}
\be
\chi_1-\chi_0\approx \alpha \frac{g^2}{\Delta^2}.
\ee
Detecting the qubit state is equivalent to detecting the cavity pull within time $T_1$.  A crude figure of merit for our ability to do this is the phase accumulation
\be
\phi = [\chi_1-\chi_0]T_1.
\ee
The qubit decay rate due to the Purcell effect is
\be
\gamma_\kappa \approx \kappa \frac{g^2}{\Delta^2}.
\ee
Assuming that there is no intrinsic (non-Purcell) relaxation (i.e.\  that the decay is Purcell dominated), then $T_1\sim 1/\gamma_\kappa$ and we have
\be
\phi\sim \frac{\alpha}{\kappa}\sim \frac{\Ec}{\kappa}
\ee
which, remarkably, is independent of $g$ and $\Delta$.  The problem is that moving the qubit closer to the cavity to enhance the homodyne signal shortens the lifetime of the qubit and reduces the allowable power in the readout beam so that the SNR does not improve.

While $\phi$ can in principle be made large, we want $\Ec$ to remain small to minimize the dephasing due to charge dispersion and we want $\kappa$ to be large so that the readout is fast compared to the intrinsic (non-Purcell) relaxation (which has been neglected in the above).  Hence it would be better to have a larger anharmonicity.  The fluxonium qubit described below will prove useful in this regard.

A more sophisticated version of the above argument that takes into account the fact that phase resolution improves with drive power is as follows.  The signal to noise power ratio for the readout to lowest order in $\chi$ is given for a (single-sided) cavity by \cite{Clerk2008}
\be
{\rm SNR} \sim 4\left[\frac{\chi_1-\chi_0}{\kappa}\right]^2 T_1 {\bar n} \kappa,
\ee
where $\bar n$ is the mean photon number in the readout cavity.  Assuming the photon number is kept at the critical value beyond which the lowest order dispersive approximation becomes invalid \cite{Gambetta2006} ($\bar n \sim \frac{\Delta^2}{4g^2}$) we see that
\be
{\rm SNR} \sim \frac{\alpha^2}{\kappa^2}\sim \frac{\Ec^2}{\kappa^2},
\ee
independent of $g$ and $\Delta$ (for $g/\Delta$ small).

The large islands of the transmon qubit make them susceptible to quasiparticle poisoning, but the very small charge dispersion means that this is almost certainly not a significant source of dephasing, though in principle quasiparticle tunneling can contribute to relaxation \cite{Lutchyn2005,Lutchyn2006,Schreier2008,MartinisQP2009,MartinisQP2009a}.  The contribution to the relaxation rate by spontaneous photon emission via the Purcell effect for charge qubits is now well-understood both theoretically and experimentally \cite{HouckPurcell2008}, but residual sources of relaxation due to dielectric losses \cite{Martinis2005,HouckPurcell2008} in the substrate or the Josephson junction oxide remain less well understood.  Small-junction charge qubits made with minimal fabrication methods seem to suffer less \cite{Schreier2008} from materials problems such as the presence of two-level glassy fluctuators \cite{Simmonds2004,Cooper2004,Faoro2006,Faoro2007,Neeley2008,Constantin2009}.

\subsection{Fluxonium Qubits}

A new qubit design \cite{fluxonium} dubbed the `fluxonium' has the closed-loop topology of the flux qubit, but the loop contains a Josephson junction series array which gives an inductance much larger than the simple geometric inductance.  This loop shorts out stray low frequency electric fields, but at the qubit frequency has such a large reactance that it is effectively an open circuit reminiscent of the charge qubit.  Initial results suggest that this qubit design is indeed robust and stable against charge noise and exhibits very long phase coherence times reaching several microseconds.  It also has the advantage that the anharmonicity can be large. Interestingly, the state-dependent polarizability and the excitation spectrum of this design is such that the qubit state can be measured even when the qubit transition frequency is driven down to very low values of order 0.5 GHz.

\section{Recent Progress and Future Directions}

Progress in the field of superconducting qubits continues at an amazing pace both in terms of fundamental non-linear quantum optics and in terms of quantum information processing.  Since the first experiments ten years ago, phase coherence times have risen approximately exponentially from immeasurably small (1 nanosecond or less) to several microseconds.  This has been achieved through a great deal of hard work and clever quantum engineering of artificial atoms and circuits by many groups. Building on this progress, DiCarlo \etal \cite{DiCarlo2009} recently demonstrated the first quantum algorithms on a two-qubit superconducting quantum processor.  Future improved qubit designs, microwave circuit designs, and materials improvements should allow this trend to continue unabated.  One interesting possibility to extend the coherence time of quantum circuits would be to use high $Q$ resonators as quantum memories or to form cavity-stabilized qubits \cite{Koch2005,Koch2006}.  Another potentially important future direction will be
hybrid systems using superconducting resonators and trapped atoms, molecules or ions \cite{Andre2006,Schustermolionstrap2009}.

A diverse and remarkable set of methods for reading out charge qubits has been developed, beginning with the charge state (dipole moment) \cite{Nakamura1999,Aassime2001} and moving on to dispersive readout of the inductive \cite{Vion2002} and capacitive \cite{BlaisCQEDtheory2004,WallraffCQED2004} susceptibilities and the low frequency readout of the quantum capacitance \cite{Duty2005,Sillanpaa2005}.  Protocols for optimal readout using linear dispersive detection in the presence of a finite energy relaxation rate for the qubit have been developed \cite{Gambetta2007}.  A novel fast latching dispersive readout using the bifurcation of a driven non-linear oscillator has proved very effective \cite{Siddiqi2006,Metcalfe2007,Boulant2007,Mallet2009} and fidelities as high as 70\% have been achieved.  The Saclay group \cite{Mallet2009} has recently reported fidelities greater than 90\% using a cavity bifurcation amplifier.  Thus, fidelities for the (approximately) QND readout \cite{Boulant2007,Lupascu2007} of charge qubits are now approaching those obtained for flux qubits and for the destructive readout of phase qubits. The measurement back action (dephasing) required \cite{Clerk2008} by the Heisenberg uncertainty principle is now understood to arise from quantum fluctuations of the light shift \cite{Schuster2005,Schuster2007a,Ithier2005,Lupascu2005} and has been observed.  The fact that this back action can be partially destructive and not simply dephasing has been understood in terms of the dressed dephasing model \cite{Boissonneault2008,Boissonneault2009}, although further quantitative experimental studies of this mechanism still need to be performed.  The Santa Barbara group has observed the non-unitary measurement back action in a `partial collapse' experiment \cite{Katz2006} in a phase qubit.  All these results demonstrate that fundamental aspects of quantum measurement theory are now accessible to condensed matter experiment for the first time.
An important future goal for QND measurements is to see quantum jumps of a qubit and observe the Zeno effect induced by continuous observation of the qubit \cite{Gambetta2008}.

It should be noted that while single-shot high fidelity readouts are extremely useful, they are in principle not essential for the exponential speed-up promised by quantum computation.  It is possible to design a sequence of quantum algorithms in which one reads out only a single bit of the $N$-bit answer with each run.  Hence, even if one only has the capability to read out a single qubit and that only with low fidelity $F$, the extra computational cost  ($\sim N/F$) is only linear.  (Of course in practice this may be a huge factor, but it is not exponential.)
High fidelity multi-qubit readout with minimal cross talk is however essential for Bell inequality tests \cite{Steffen2006a} that close the detector loophole.  Using a co-planar waveguide resonator rather than capacitive coupling, the Santa Barbara group has recently greatly reduced the cross talk of their readout scheme for phase qubits and closed the measurement loophole for the Bell violation \cite{UCSB_Bell2009}.

An important idea in circuit QED is the understanding that dispersive coupling to the cavity can be used to perform a simultaneous joint readout of multiple qubits \cite{BlaisCQEDtheory2004,Bishop2009a}.  This joint readout was used for two-qubit state tomography by Majer \etal \cite{Majer2007}. In its most basic form, the idea is simply that with two qubits, there are four possible quantum states and four different dispersive frequency pulls of the cavity.  If one is in the strong dispersive coupling regime and all four frequency pulls can be reliably distinguished in a single shot, then one has two bits of classical information and a complete projective measurement of both qubits. Of course in the presence of qubit decay and amplifier noise, the detector tomography can be complex \cite{Bishop2009a}.  A more sophisticated understanding of the situation of imperfect resolution of the four peaks has been developed recently and Filipp \etal \cite{Filipp2009} demonstrated that it is possible to still reliably measure two-qubit correlations even in the presence of readout noise.

 A simplified version of the theory for joint readout of two qubits is the following.  While the cavity pull is linear in the qubit polarizations
\be
\delta\omega_{\rm c} = \chi_1 \sigma_1^z + \chi_2 \sigma_2^z,
\ee
the corresponding homodyne (transmission) amplitude is \emph{not}
\be
A(\sigma_1^z,\sigma_2^z) = \mathrm{Re}\left\{e^{i\varphi} \frac{\kappa/2}{\Delta - \delta\omega_{\rm c}+i\kappa/2}   \right\}.
\ee
Here, $\Delta$ is the detuning of the readout tone from the bare cavity resonance, $\kappa$ is the cavity line width, and $\varphi$ is the local oscillator phase.  Because this can take on only four distinct values (corresponding to two classical bits of information) this expression can always be recast in the form
\be
A(\sigma_1^z,\sigma_2^z)=\beta_0 + \beta_1 \sigma_1^z + \beta_2 \sigma_2^z + \beta_{12} \sigma_1^z\sigma_2^z.
\ee
The joint coefficient $\beta_{12}$ is in general non-zero (as long as $\Delta\ne 0$) and typically on the same scale as the other coefficients.  By using pre-rotations (by angle zero or $\pi$) of each of the two qubits prior to making the measurement, it is straightforward to obtain any one or two-qubit correlator in the $z$ basis.  Ensemble averaging many such measurements will reduce the statistical uncertainty to arbitrarily low values. For example,
\be
\langle \sigma_1^z\sigma_2^z\rangle = \frac{1}{4\beta_{12}}\left\langle A(\sigma_1^z,\sigma_2^z)-A(-\sigma_1^z,\sigma_2^z)-A(\sigma_1^z,-\sigma_2^z)+A(\sigma_1^z,\sigma_2^z)\right\rangle.
\ee
Any other arbitrary correlators (e.g. $\langle \sigma_1^x \sigma_2^y\rangle$) can be achieved by pre-pending rotations through appropriate arbitrary angles.  The Yale group has recently used this to measure values of the CHSH entanglement witness well above the classical bound \cite{EntanglementMetrology2009}.

High-fidelity single qubit gates \cite{Wallraff2005,Lucero2008} have been developed allowing observation of the Berry phase for spin 1/2 \cite{Leek2007} using a charge qubit and higher spins \cite{Neeley2009} simulated using the multilevel structure of a phase qubit.  This is a first step towards larger scale quantum simulators. State and process tomography for one and two qubit operations \cite{Steffen2006,Steffen2006a,Majer2007} and randomized benchmarking \cite{Chow2009} are now routine.  A number of methods for two-qubit gates have been suggested \cite{Blais03,Rigetti2005,Bertet2006,Gywat2006,Blais2007,Sillanpaa2007,Plantenberg2007} involving fixed capacitive couplings \cite{Pashkin2003,Yamamoto2003,Berkley2003,McDermott2005}, inductive coupling \cite{Majer2005}, and virtual photon exchange via a cavity bus \cite{Majer2007,Sillanpaa2007}.  Controlled phase gates have been proposed \cite{Strauch2003,Ke-Hui2007} and realized using virtual states outside the logical basis \cite{DiCarlo2009}.    A key problem for the future is to further increase the on-off ratio of controllable qubit couplings.  One possible resource for this would be tunable cavities \cite{Wallquist2006,Palacios-Laloy2008,Sandberg2008} or tunable couplings via an active element \cite{Hime2006,Niskanen2007}.  It seems clear however that simply tuning a coupling element off resonance will not produce an adequately large on-off ratio.  One may be able to do better by using a tunable interference between two coherent coupling channels to actually null out the coupling at a special operating point \cite{SMGRJSunpublished}.

Given the current rate of progress, it will not be long before other two qubit algorithms \cite{DobsicekPhaseEstimation2007} and quantum information processing with more than two superconducting qubits will be realized.  Key short term goals will be to create multi-qubit entangled states such as the GHZ and W states \cite{Wei2006,Tsomokos2008,Bishop2009a,Galiautdinov2008}, and begin to execute simple error correction protocols \cite{MikeandIke,TornbergPhaseFlip2008}.

Another exciting direction involves using multiple physical qubits to realize individual logical qubits to overcome the difficulties of maintaining stable transition frequencies.  In particular, the possibility of topological protection \cite{Kitaev2003,Ioffe2002,Ioffe2002a,Doucot2003,Doucot2005} is beginning to be explored in superconducting qubits \cite{Gladchenko2009}.  The central idea is that qubits are constructed in which the ground and excited states are degenerate and this degeneracy is robust against local variations in Hamiltonian parameters.  Even if the energy levels are not exactly degenerate, it would be very useful to have a qubit with a ``Lambda" energy level scheme, that is, two nearly degenerate levels that can be coupled via stimulated Raman pulses through a third level.  This would be advantageous both as a robust qubit and for purposes of fundamental quantum optics studies.  It seems reasonably certain that this cannot be achieved without applied magnetic flux to frustrate the Josephson couplings (as in a flux qubit or in the fluxonium qubit).  Indeed the fluxonium qubit may turn out to be quite useful as a Lambda system.

\begin{figure}[htp]
\centering
\includegraphics[totalheight=0.3\textheight,clip]{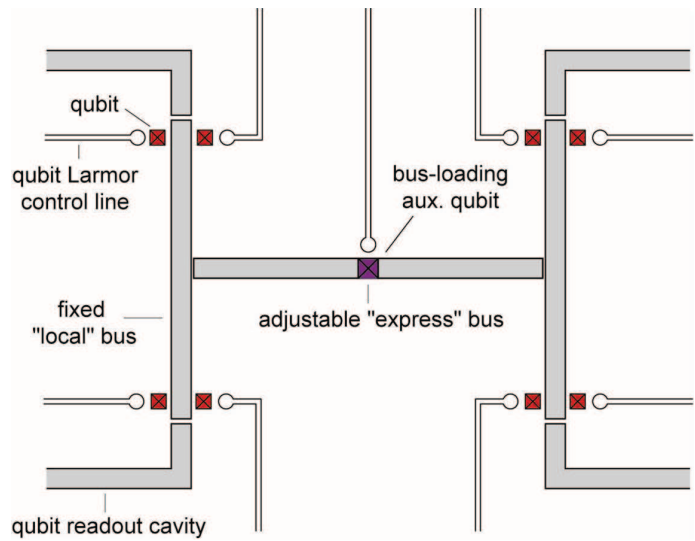}
\caption{Circuit QED: Schematic illustration of a possible architecture for an eight-qubit processor.  Four transmon qubits are embedded in each quantum bus (a coplanar waveguide resonator). The two resonators in turn are connected by an `express bus' consisting of a tunable resonator whose resonance frequency can be rapidly moved.}\label{fig:8_qubit_processor}
\end{figure}

To scale up to more qubits in the circuit QED scheme, it will be necessary to move to two cavities \cite{Ogden2008}
and ultimately to cavity grids \cite{Helmer2009}.  A possible architecture for an eight-qubit processor is shown in figure \ref{fig:8_qubit_processor}.

The case of large cavity arrays will be interesting not only as a quantum computation architecture but also for fundamental quantum optics purposes.  An array of resonators each containing a qubit that induces a Kerr nonlinearity will be a realization of the boson Hubbard model \cite{Fisher1989} which exhibits both superfluid and Mott insulator phases.  There is now a burgeoning interest in seeing `quantum phase transitions of light' \cite{Greentree2006,Illuminati2006,Hartmann2007,Jarrett2007,Rossini2007,Hartmann2008,Makin2008,
Aichhorn2008,Cho2008,Cho2008a,Zhao2008,Na2008,Lei2008,
Ji2009,LaserPhotonicsReview2009,Grochol2009,Dalidovich2009,Carusotto2009,
Schmidt2009,Koch2009}.  Since the transmon qubit is itself an anharmonic oscillator, one might imagine it would be easier to simply use a  lattice of coupled transmons to realize the boson Hubbard model (with negative Kerr coefficient). The advantage of using a lattice of resonators is that their resonance frequencies can be closely matched to a single fixed value.  The Kerr coefficient induced by coupling each resonator to an off-resonant qubit will have some variation due to variations in qubit transition frequencies, but this disorder in the Hubbard $U$ will be more tolerable than disorder in the photon `site energies.' Just as cold atom systems are now used to simulate condensed matter models, so we may be able to use photons as  interacting strongly correlated bosons, which can be probed, measured and controlled in ways that are impossible in ordinary condensed matter.

In summary, the future of circuit QED looks bright indeed for both practical applications (such as quantum limited amplifiers) and fundamental new physics with artificial atoms and microwave photons.   Circuit QED is much more than atomic physics with wires.  We have a set of modular elements which are readily connected together (spatial mode matching is easy with wires!). Hence, we have the opportunity to assemble large scale structures from these quantum building blocks and do some real quantum engineering.  While our cold atoms colleagues are busy trying to emulate condensed matter systems, we may be able to use topological quantum computation and quantum error correction schemes to realize non-abelian gauge theories of particle physics.

\ack{The authors are grateful for research support provided by Yale University, the National Science Foundation, the Army Research Office, and IARPA.  They are also grateful to the numerous colleagues with whom they have had enlightening discussions and collaborations on the quantum mechanics of electrical circuits.}
\section*{References}
\bibliographystyle{unsrt}
%
\bibliography{NobelYaleBibFile_2}

\begin{thebibliography}{100}

\bibitem{Devoret2004}
MH~Devoret and JM~Martinis.
\newblock {Superconducting qubits}.
\newblock In {Esteve, D and Raimond, JM and Dalibard, J}, editor, {\em {Quantum
  Entanglement and Information Processing}}, volume~{79} of {\em {LES HOUCHES
  SUMMER SCHOOL SESSION}}, pages {443--485}, {2004}.
\newblock {Les Houches Session 79th on Quantum Entanglement and Information
  Processing, Les Houches, FRANCE, JUN 30-JUL 25, 2003}.

\bibitem{Esteve2005}
D~Esteve and D~Vion.
\newblock {Solid state quantum bit circuits}.
\newblock In {Bouchiat, H and Gefen, Y and Gueron, S and Montambaux, G and
  Dalibard, J}, editor, {\em {Nanophysics: Coherence and Transport}},
  volume~{81} of {\em {LES HOUCHES SUMMER SCHOOL SESSION}}, pages {537+},
  {2005}.
\newblock {Les Houches Session 81st on Nanophysics - Coherence and Transport,
  Les Houches, FRANCE, JUN 28-JUL 30, 2004}.

\bibitem{Wendin_Shumeiko2006}
G.~Wendin and V.S. Shumeiko.
\newblock {\em Handbook of Theoretical and Computational Nanotechnology, Vol.
  3}, chapter `Superconducting circuits, qubits and computing', pages 223--309.
\newblock American Scientific Publishers, Los Angeles, 2006.

\bibitem{Wendin_Shumeiko2007}
G.~Wendin and V.~S. Shumeiko.
\newblock Quantum bits with josephson junctions.
\newblock {\em J. Low Temperature Physics}, 33:724, 2007.

\bibitem{Clarke2008}
John Clarke and Frank~K. Wilhelm.
\newblock {Superconducting quantum bits}.
\newblock {\em Nature}, {453}:{1031--1042}, {2008}.

\bibitem{WiringUpQuantumSystems}
RJ~Schoelkopf and SM~Girvin.
\newblock Wiring up quantum systems.
\newblock {\em Nature}, 451:664, 2008.

\bibitem{YouandNori2005}
JQ~You and F~Nori.
\newblock {Superconducting circuits and quantum information}.
\newblock {\em Physics Today}, {58}:{42--47}, {2005}.

\bibitem{Nori2008}
Franco Nori.
\newblock Superconducting qubits: Atomic physics with a circuit.
\newblock {\em Nature Physics}, 4:589--590, 2008.

\bibitem{Korotkov2009}
Alexander~N. Korotkov.
\newblock {Special issue on quantum computing with superconducting qubits}.
\newblock {\em Quantum Information Processing}, {8}:{51--54}, {2009}.

\bibitem{Leggett1980}
AJ~Leggett.
\newblock Macroscopic quantum systems and the quantum theory of measurement.
\newblock {\em Progress of Theoretical Physics Suppl.}, 69:80--100, 1980.

\bibitem{MartinisDevoretJJreview}
JM~Martinis, A~Wallraff, and MH~Devoret.
\newblock Superconducting qubits: A short review.
\newblock {\em cond-mat/0411.174}, 2004.

\bibitem{MartinisDevoretClarke1985}
JM~Martinis, MH~Devoret, and J~Clarke.
\newblock Energy level quantization in the zero-voltage state of a
  current-biased josephson junction.
\newblock {\em Phys. Rev. Lett.}, 55:1543 -- 1546, 1985.

\bibitem{Clarke1988Science}
John Clarke, Andrew~N. Cleland, Michel~H. Devoret, Daniel Esteve, and John~M.
  Martinis.
\newblock {Quantum Mechanics of a Macroscopic Variable: The Phase Difference of
  a Josephson Junction}.
\newblock {\em Science}, 239:992--997, 1988.

\bibitem{CaldeiraandLeggett1983}
AO~Caldeira and AJ~Leggett.
\newblock Quantum tunneling in a dissipative system.
\newblock {\em Annals of Physics}, 149:374--456, 1983.

\bibitem{VossandWebb1981}
Richard~F Voss and Richard~A Webb.
\newblock Macroscopic quantum tunneling in 1 $\mu$-m nb josephson junctions.
\newblock {\em Phys. Rev. Lett.}, 47:265 -- 268, 1981.

\bibitem{DevoretMartinisClarkeMQT1985}
Michel~H Devoret, Martinis, John M, and John Clarke.
\newblock Measurements of macroscopic quantum tunneling out of the zero-voltage
  state of a current-biased josephson junction.
\newblock {\em Phys. Rev. Lett.}, 55:1908 -- 1911, 1985.

\bibitem{Turlot1989}
Emmanuel Turlot, Daniel Esteve, Cristian Urbina, John~M. Martinis, Michel~H.
  Devoret, Sebastian Linkwitz, and Hermann Grabert.
\newblock Escape oscillations of a josephson junction switching out of the
  zero-voltage state.
\newblock {\em Phys. Rev. Lett.}, 62:1788--1791, 1989.

\bibitem{DevoretPofE}
M.~H. Devoret, D.~Esteve, H.~Grabert, G.-L. Ingold, H.~Pothier, and C.~Urbina.
\newblock Effect of the electromagnetic environment on the coulomb blockade in
  ultrasmall tunnel junctions.
\newblock {\em Phys. Rev. Lett.}, 64:1824--1827, 1990.

\bibitem{GirvinPofE}
S.~M. Girvin, L.~I. Glazman, M.~Jonson, D.~R. Penn, and M.~D. Stiles.
\newblock Quantum fluctuations and the single-junction coulomb blockade.
\newblock {\em Phys. Rev. Lett.}, 64:3183--3186, 1990.

\bibitem{Purcell1946}
E.~M. Purcell.
\newblock Spontaneous emission probabilities at radio frequencies.
\newblock {\em Phys. Rev.}, 69:681, 1946.

\bibitem{HouckPurcell2008}
A.~A. Houck, J.~A. Schreier, B.~R. Johnson, J.~M. Chow, Jens Koch, J.~M.
  Gambetta, D.~I. Schuster, L.~Frunzio, M.~H. Devoret, S.~M. Girvin, and R.~J.
  Schoelkopf.
\newblock {Controlling the spontaneous emission of a superconducting transmon
  qubit}.
\newblock {\em Phys. Rev. Lett.}, {101}:080502, {2008}.

\bibitem{Houck2007SinglePhoton}
A.~A. Houck, D.~I. Schuster, J.~M. Gambetta, J.~A. Schreier, B.~R. Johnson,
  J.~M. Chow, L.~Frunzio, J.~Majer, M.~H. Devoret, S.~M. Girvin, and R.~J.
  Schoelkopf.
\newblock {Generating single microwave photons in a circuit}.
\newblock {\em Nature}, {449}:{328--331}, {2007}.

\bibitem{Neeley2008a}
Matthew Neeley, M.~Ansmann, Radoslaw~C. Bialczak, M.~Hofheinz, N.~Katz, Erik
  Lucero, A.~O'Connell, H.~Wang, A.~N. Cleland, and John~M. Martinis.
\newblock Transformed dissipation in superconducting quantum circuits.
\newblock {\em Phys. Rev. B}, 77:180508, 2008.

\bibitem{Clerk2008}
A.~A. Clerk, M.~H. Devoret, S.~M. Girvin, F.~Marquardt, and R.~J. Schoelkopf.
\newblock Introduction to quantum noise, measurement and amplification.
\newblock {\em arXiv.org:0810.4729}, 2008.
\newblock (Rev. Mod. Phys., in press).

\bibitem{Tavis1968}
Michael Tavis and Frederick~W. Cummings.
\newblock Exact solution for an $n$-molecule radiation-field hamiltonian.
\newblock {\em Phys. Rev.}, 170:379--384, 1968.

\bibitem{Mabuchi02}
H.~Mabuchi and A.~Doherty.
\newblock Cavity quantum electrodynamics: Coherence in context.
\newblock {\em Science}, 298:1372, 2002.

\bibitem{Walls94}
D.~Walls and G.~Milburn.
\newblock {\em Quantum optics}.
\newblock Spinger-Verlag, Berlin, 1994.

\bibitem{Thompson1992}
R.~J. Thompson, G.~Rempe, and H.~J. Kimble.
\newblock Observation of normal-mode splitting for an atom in an optical
  cavity.
\newblock {\em Phys. Rev. Lett.}, 68:1132--1135, 1992.

\bibitem{Boca2004}
A~Boca, R~Miller, KM~Birnbaum, AD~Boozer, J~McKeever, and HJ~Kimble.
\newblock {Observation of the vacuum Rabi spectrum for one trapped atom}.
\newblock {\em Phys. Rev. Lett.}, {93}:233603, {2004}.

\bibitem{SchusterAMOVacRabi2008}
I.~Schuster, A.~Kubanek, A.~Fuhrmanek, T.~Puppe, P.~W.~H. Pinkse, K.~Murr, and
  G.~Rempe.
\newblock {Nonlinear spectroscopy of photons bound to one atom}.
\newblock {\em Nature Physics}, {4}:{382--385}, {2008}.

\bibitem{Nogues1999}
G.~Nogues, A.~Rauschenbeutel, S.~Osnaghi, M.~Brune, J.~M. Raimond, and
  S.~Haroche.
\newblock Seeing a single photon without destroying it.
\newblock {\em Nature}, 400:239--242, 1999.

\bibitem{Guerlin2007}
Christine Guerlin, Julien Bernu, Samuel Deleglise, Clement Sayrin, Sebastien
  Gleyzes, Stefan Kuhr, Michel Brune, Jean-Michel Raimond, and Serge Haroche.
\newblock Progressive field-state collapse and quantum non-demolition photon
  counting.
\newblock {\em Nature}, 448:889--893, 2007.

\bibitem{Gleyzes2007}
Sebastien Gleyzes, Stefan Kuhr, Christine Guerlin, Julien Bernu, Samuel
  Deleglise, Ulrich Busk~Hoff, Michel Brune, Jean-Michel Raimond, and Serge
  Haroche.
\newblock Quantum jumps of light recording the birth and death of a photon in a
  cavity.
\newblock {\em Nature}, 446:297--300, 2007.

\bibitem{Deleglise2008}
Samuel Deleglise, Igor Dotsenko, Clement Sayrin, Julien Bernu, Michel Brune,
  Jean-Michel Raimond, and Serge Haroche.
\newblock Reconstruction of non-classical cavity field states with snapshots of
  their decoherence.
\newblock {\em Nature}, 455:510--514, 2008.

\bibitem{HarocheRaimondRMP}
J.~Raimond, M.~Brune, and S.~Haroche.
\newblock Manipulating quantum entanglement with atoms and photons in a cavity.
\newblock {\em Rev. Mod. Phys.}, 73:565, 2001.

\bibitem{HarocheRaimondcQEDBook}
Serge Haroche and Jean-Michel Raimond.
\newblock {\em Exploring the Quantum: Atoms, Cavities and Photons}.
\newblock Oxford University Press, 2006.

\bibitem{BlaisCQEDtheory2004}
Alexandre Blais, Ren-Shou Huang, Andreas Wallraff, S.~M. Girvin, and R.~J.
  Schoelkopf.
\newblock Cavity quantum electrodynamics for superconducting electrical
  circuits: an architecture for quantum computation.
\newblock {\em Phys. Rev. A}, 69:062320, 2004.

\bibitem{WallraffCQED2004}
A.~Wallraff, D.~I. Schuster, A.~Blais, L.~Frunzio, R.-S. Huang, J.~Majer,
  S.~Kumar, S.~M. Girvin, and R.~J. Schoelkopf.
\newblock Circuit quantum electrodynamics: Coherent coupling of a single photon
  to a cooper pair box.
\newblock {\em Nature}, 431:162--167, 2004.

\bibitem{Chiorescu2004}
I.~Chiorescu, P.~Bertet, K.~Semba, Y.~Nakamura, C.~J. P.~M. Harmans, and J.~E.
  Mooij.
\newblock Coherent dynamics of a flux qubit coupled to a harmonic oscillator.
\newblock {\em Nature}, 431:159--162, 2004.

\bibitem{Devoret2007}
Michel Devoret, Steven Girvin, and Robert Schoelkopf.
\newblock {Circuit-QED: How strong can the coupling between a Josephson
  junction atom and a transmission line resonator be?}
\newblock {\em Annalen der Physik}, {16}:{767--779}, {2007}.

\bibitem{Shnirman97}
A.~Shnirman, G.~Sch\"on, and Z.~Hermon.
\newblock Quantum manipulations of small josephson junctions.
\newblock {\em Phys. Rev. Lett.}, 79:2371, 1997.

\bibitem{Makhlin01}
Y.~Makhlin, G.~Sch\"on, and A.~Shnirman.
\newblock Quantum-state engineering with josephson-junction devices.
\newblock {\em Rev. Mod. Phys.}, 73:357, 2001.

\bibitem{Buisson01}
O.~Buisson and F.~Hekking.
\newblock {\em Macroscopic Quantum Coherence and Quantum Computing}, chapter
  Entangled states in a Josephson charge qubit coupled to a superconducting
  resonator.
\newblock Kluwer, New York, 2001.

\bibitem{Marquardt01}
F.~Marquardt and C.~Bruder.
\newblock Superposition of two mesoscopically distinct quantum states: Coupling
  a cooper pair box to a large superconducting island.
\newblock {\em Phys. Rev. B}, 63:054514, 2001.

\bibitem{Al-Saidi01}
W.~A. Al-Saidi and D.~Stroud.
\newblock Eigenstates of a small josephson junction coupled to a resonant
  cavity.
\newblock {\em Phys. Rev. B}, 65:014512, 2001.

\bibitem{Plastina03}
F.~Plastina and G.~Falci.
\newblock Communicating josephson qubits.
\newblock {\em Phys. Rev. B}, 67:224514, 2003.

\bibitem{Blais03}
A.~Blais, A.~Maassen van~den Brink, and A.~Zagoskin.
\newblock Tunable coupling of superconducting qubits.
\newblock {\em Phys. Rev. Lett.}, 90:127901, 2003.

\bibitem{Yang03}
C.-P. Yang, S.-I. Chu, and S.~Han.
\newblock Possible realization of entanglement, logical gates, and
  quantum-information transfer with superconducting-quantum-interference device
  qubits in cavity qed.
\newblock {\em Phys. Rev. A}, 67:042311, 2003.

\bibitem{you03a}
J.~Q. You and F.~Nori.
\newblock Quantum information processing with superconducting qubits in a
  microwave field.
\newblock {\em Phys. Rev. B}, 68:064509, 2003.

\bibitem{Schuster2005}
DI~Schuster, A~Wallraff, A~Blais, L~Frunzio, RS~Huang, J~Majer, SM~Girvin, and
  RJ~Schoelkopf.
\newblock {ac Stark shift and dephasing of a superconducting qubit strongly
  coupled to a cavity field}.
\newblock {\em Phys. Rev. Lett.}, {94}:123602, {2005}.

\bibitem{Wallraff2005}
A~Wallraff, DI~Schuster, A~Blais, L~Frunzio, J~Majer, MH~Devoret, SM~Girvin,
  and RJ~Schoelkopf.
\newblock {Approaching unit visibility for control of a superconducting qubit
  with dispersive readout}.
\newblock {\em Phys. Rev. Lett.}, {95}:060501, {2005}.

\bibitem{Schuster2007a}
D.~I. Schuster, A.~Wallraff, A.~Blais, L.~Frunzio, R.-S. Huang, J.~Majer, S.~M.
  Girvin, and R.~J. Schoelkopf.
\newblock Erratum: ac stark shift and dephasing of a superconducting qubit
  strongly coupled to a cavity field {[\emph{Phys. Rev. Lett.} {\bf 94}, 123602
  (2005)]}.
\newblock {\em Phys. Rev. Lett.}, 98:049902, 2007.

\bibitem{Johansson2006}
J.~Johansson, S.~Saito, T.~Meno, H.~Nakano, M.~Ueda, K.~Semba, and
  H.~Takayanagi.
\newblock Vacuum rabi oscillations in a macroscopic superconducting qubit lc
  oscillator system.
\newblock {\em Phys. Rev. Lett.}, 96:127006, 2006.

\bibitem{Siddiqi2006}
I~Siddiqi, R~Vijay, M~Metcalfe, E~Boaknin, L~Frunzio, RJ~Schoelkopf, and
  MH~Devoret.
\newblock {Dispersive measurements of superconducting qubit coherence with a
  fast latching readout}.
\newblock {\em Phys. Rev. B}, {73}:054510, {2006}.

\bibitem{Boulant2007}
N.~Boulant, G.~Ithier, P.~Meeson, F.~Nguyen, D.~Vion, D.~Esteve, I.~Siddiqi,
  R.~Vijay, C.~Rigetti, F.~Pierre, and M.~Devoret.
\newblock Quantum nondemolition readout using a josephson bifurcation
  amplifier.
\newblock {\em Phys. Rev. B}, 76:014525, 2007.

\bibitem{Schuster2007}
D.I. Schuster, A.A. Houck, J.A. Schreier, A.~Wallraff, J.~Gambetta, A.~Blais,
  L.~Frunzio, B.~Johnson, M.H. Devoret, S.M. Girvin, and R.J. Schoelkopf.
\newblock Resolving photon number states in a superconducting circuit.
\newblock {\em Nature}, 445:515--518, 2007.

\bibitem{Wallraff2007}
A.~Wallraff, D.~I. Schuster, A.~Blais, J.~M. Gambetta, J.~Schreier, L.~Frunzio,
  M.~H. Devoret, S.~M. Girvin, and R.~J. Schoelkopf.
\newblock {Sideband transitions and two-tone spectroscopy of a superconducting
  qubit strongly coupled to an on-chip cavity}.
\newblock {\em Phys. Rev. Lett.}, {99}:050501, {2007}.

\bibitem{Sillanpaa2007}
Mika~A. Sillanp\"a\"a, Jae~I. Park, and Raymond~W. Simmonds.
\newblock {Coherent quantum state storage and transfer between two phase qubits
  via a resonant cavity}.
\newblock {\em Nature}, {449}:{438--442}, {2007}.

\bibitem{Leek2007}
P.~J. Leek, J.~M. Fink, A.~Blais, R.~Bianchetti, M.~Goeppl, J.~M. Gambetta,
  D.~I. Schuster, L.~Frunzio, R.~J. Schoelkopf, and A.~Wallraff.
\newblock {Observation of Berry's phase in a solid-state qubit}.
\newblock {\em Science}, {318}:{1889--1892}, {2007}.

\bibitem{Majer2007}
J.~Majer, J.~M. Chow, J.~M. Gambetta, Jens Koch, B.~R. Johnson, J.~A. Schreier,
  L.~Frunzio, D.~I. Schuster, A.~A. Houck, A.~Wallraff, A.~Blais, M.~H.
  Devoret, S.~M. Girvin, and R.~J. Schoelkopf.
\newblock {Coupling superconducting qubits via a cavity bus}.
\newblock {\em Nature}, {449}:{443--447}, {2007}.

\bibitem{Astafiev2007}
O.~Astafiev, K.~Inomata, A.~O. Niskanen, T.~Yamamoto, Yu.~A. Pashkin,
  Y.~Nakamura, and J.~S. Tsai.
\newblock {Single artificial-atom lasing}.
\newblock {\em Nature}, {449}:{588--590}, {2007}.

\bibitem{Metcalfe2007}
M.~Metcalfe, E.~Boaknin, V.~Manucharyan, R.~Vijay, I.~Siddiqi, C.~Rigetti,
  L.~Frunzio, R.~J. Schoelkopf, and M.~H. Devoret.
\newblock Measuring the decoherence of a quantronium qubit with the cavity
  bifurcation amplifier.
\newblock {\em Phys. Rev. B}, 76:174516, 2007.

\bibitem{Deppe2008}
Frank Deppe, Matteo Mariantoni, E.~P. Menzel, A.~Marx, S.~Saito, K.~Kakuyanagi,
  H.~Tanaka, T.~Meno, K.~Semba, H.~Takayanagi, E.~Solano, and R.~Gross.
\newblock {Two-photon probe of the Jaynes-Cummings model and controlled
  symmetry breaking in circuit QED}.
\newblock {\em Nature Physics}, {4}:{686--691}, {2008}.

\bibitem{Fink2008}
J.~M. Fink, M.~Goeppl, M.~Baur, R.~Bianchetti, P.~J. Leek, A.~Blais, and
  A.~Wallraff.
\newblock {Climbing the Jaynes-Cummings ladder and observing its root n
  nonlinearity in a cavity QED system}.
\newblock {\em Nature}, {454}:{315--318}, {2008}.

\bibitem{Hofheinz2008}
Max Hofheinz, E.~M. Weig, M.~Ansmann, Radoslaw~C. Bialczak, Erik Lucero,
  M.~Neeley, A.~D. O'Connell, H.~Wang, John~M. Martinis, and A.~N. Cleland.
\newblock {Generation of Fock states in a superconducting quantum circuit}.
\newblock {\em Nature}, {454}:{310--314}, {2008}.

\bibitem{Wang2008}
H.~Wang, M.~Hofheinz, M.~Ansmann, R.~C. Bialczak, E.~Lucero, M.~Neeley, A.~D.
  O'Connell, D.~Sank, J.~Wenner, A.~N. Cleland, and John~M. Martinis.
\newblock {Measurement of the Decay of Fock States in a Superconducting Quantum
  Circuit}.
\newblock {\em Phys. Rev. Lett.}, {101}:240401, {2008}.

\bibitem{Schreier2008}
J.~A. Schreier, A.~A. Houck, Jens Koch, D.~I. Schuster, B.~R. Johnson, J.~M.
  Chow, J.~M. Gambetta, J.~Majer, L.~Frunzio, M.~H. Devoret, S.~M. Girvin, and
  R.~J. Schoelkopf.
\newblock {Suppressing charge noise decoherence in superconducting charge
  qubits}.
\newblock {\em Phys. Rev. B}, {77}:180502, {2008}.

\bibitem{Fragner2008}
A.~Fragner, M.~Goeppl, J.~M. Fink, M.~Baur, R.~Bianchetti, P.~J. Leek,
  A.~Blais, and A.~Wallraff.
\newblock {Resolving Vacuum Fluctuations in an Electrical Circuit by Measuring
  the Lamb Shift}.
\newblock {\em Science}, {322}:{1357--1360}, {2008}.

\bibitem{Grajcar2008}
M.~Grajcar, S.~H.~W. van~der Ploeg, A.~Izmalkov, E.~Il'ichev, H.-G. Meyer,
  A.~Fedorov, A.~Shnirman, and Gerd Schon.
\newblock Sisyphus cooling and amplification by a superconducting qubit.
\newblock {\em Nature Physics}, 4:612--616, 2008.

\bibitem{Sandberg2008}
M.~Sandberg, C.~M. Wilson, F.~Persson, G.~Johansson, V.~Shumeiko, T.~Bauch,
  T.~Duty, and P.~Delsing.
\newblock {Fast tuning of superconducting microwave cavities}.
\newblock In {Goan, HS and Chen, YN}, editor, {\em Solid-state Quantum
  Computing, Proceedings}, volume {1074} of {\em {AIP CONFERENCE PROCEEDINGS}},
  pages {12--21}, {2008}.
\newblock {2nd International Workshop on Solid-State Quantum
  Computing/Mini-School on Quantum Information Science, Taipei, TAIWAN, JUN
  23-27, 2008}.

\bibitem{Il'ichev2009}
E.~Il'ichev, S.~H.~W. van~der Ploeg, M.~Grajcar, and H.~G. Meyer.
\newblock {Weak continuous measurements of multiqubits systems}.
\newblock {\em Quantum Information Processing}, {8}:{133--153}, {2009}.

\bibitem{Bishop2009}
Lev~S. Bishop, J.~M. Chow, Jens Koch, A.~A. Houck, M.~H. Devoret, E.~Thuneberg,
  S.~M. Girvin, and R.~J. Schoelkopf.
\newblock {Nonlinear response of the vacuum Rabi resonance}.
\newblock {\em Nature Physics}, {5}:{105--109}, {2009}.

\bibitem{Chow2009}
J.~M. Chow, J.~M. Gambetta, L.~Tornberg, Jens Koch, Lev~S. Bishop, A.~A. Houck,
  B.~R. Johnson, L.~Frunzio, S.~M. Girvin, and R.~J. Schoelkopf.
\newblock {Randomized Benchmarking and Process Tomography for Gate Errors in a
  Solid-State Qubit}.
\newblock {\em Phys. Rev. Lett.}, {102}:090502, {2009}.

\bibitem{Hofheinz2009}
Max Hofheinz, H.~Wang, M.~Ansmann, Radoslaw~C. Bialczak, Erik Lucero,
  M.~Neeley, A.~D. O'Connell, D.~Sank, J.~Wenner, John~M. Martinis, and A.~N.
  Cleland.
\newblock {Synthesizing arbitrary quantum states in a superconducting
  resonator}.
\newblock {\em Nature}, {459}:{546--549}, {2009}.

\bibitem{DiCarlo2009}
L.~DiCarlo, J.~M. Chow, J.~M. Gambetta, Lev~S. Bishop, B.~R. Johnson, D.~I.
  Schuster, J.~Majer, A.~Blais, L.~Frunzio, S.~M. Girvin, and R.~J. Schoelkopf.
\newblock Demonstration of two-qubit algorithms with a superconducting quantum
  processor.
\newblock {\em Nature}, 460:240--244, 2009.

\bibitem{Koch2007}
Jens Koch, Terri~M. Yu, Jay Gambetta, A.~A. Houck, D.~I. Schuster, J.~Majer,
  Alexandre Blais, M.~H. Devoret, S.~M. Girvin, and R.~J. Schoelkopf.
\newblock {Charge-insensitive qubit design derived from the Cooper pair box}.
\newblock {\em Phys. Rev. A}, {76}:042319, {2007}.

\bibitem{Fluxqubitstrongcoupling2009}
J\'er\^ome Bourassa, Jay~M. Gambetta, Abdufarrukh A.~Abdumalikov Jr, Oleg
  Astafiev, and Alexandre~Blais Yasunobu~Nakamura.
\newblock Ultra-strong coupling regime of cavity qed with phase-biased flux
  qubits.
\newblock {\em arXiv:0906.1383}, 2009.

\bibitem{Gambetta2006}
Jay Gambetta, Alexandre Blais, D.~I. Schuster, A.~Wallraff, L.~Frunzio,
  J.~Majer, M.~H. Devoret, S.~M. Girvin, and R.~J. Schoelkopf.
\newblock {Qubit-photon interactions in a cavity: Measurement-induced dephasing
  and number splitting}.
\newblock {\em Phys. Rev. A}, {74}:042318, {2006}.

\bibitem{Bertet2005}
P.~Bertet, I.~Chiorescu, G.~Burkard, K.~Semba, C.~J. P.~M. Harmans, D.~P.
  DiVincenzo, and J.~E. Mooij.
\newblock Dephasing of a superconducting qubit induced by photon noise.
\newblock {\em Phys. Rev. Lett.}, 95:257002, 2005.

\bibitem{Ithier2005}
G~Ithier, E~Collin, P~Joyez, PJ~Meeson, D~Vion, D~Esteve, F~Chiarello,
  A~Shnirman, Y~Makhlin, J~Schriefl, and G~Schon.
\newblock {Decoherence in a superconducting quantum bit circuit}.
\newblock {\em Phys. Rev. B}, {72}:134519, {2005}.

\bibitem{Lupascu2005}
A~Lupascu, CJPM Harmans, and JE~Mooij.
\newblock {Quantum state detection of a superconducting flux qubit using a
  dc-SQUID in the inductive mode}.
\newblock {\em Phys. Rev. B}, {71}:184506, {2005}.

\bibitem{Boissonneault2008}
Maxime Boissonneault, J.~M. Gambetta, and Alexandre Blais.
\newblock {Nonlinear dispersive regime of cavity QED: The dressed dephasing
  model}.
\newblock {\em Phys. Rev. A}, {77}:060305, {2008}.

\bibitem{Boissonneault2009}
Maxime Boissonneault, J.~M. Gambetta, and Alexandre Blais.
\newblock {Dispersive regime of circuit QED: Photon-dependent qubit dephasing
  and relaxation rates}.
\newblock {\em Phys. Rev. A}, {79}:013819, {2009}.

\bibitem{Helmer2009a}
Ferdinand Helmer, Matteo Mariantoni, Enrique Solano, and Florian Marquardt.
\newblock Quantum nondemolition photon detection in circuit qed and the quantum
  zeno effect.
\newblock {\em Phys. Rev. A}, 79:052115, 2009.

\bibitem{Romero2009}
G.~Romero, J.~J. Garcia-Ripoll, and E.~Solano.
\newblock {Microwave Photon Detector in Circuit QED}.
\newblock {\em Phys. Rev. Lett.}, {102}:173602, {2009}.

\bibitem{Marthaler2008}
M.~Marthaler, Gerd Sch\"on, and Alexander Shnirman.
\newblock {Photon-number squeezing in circuit quantum electrodynamics}.
\newblock {\em Phys. Rev. Lett.}, {101}:147001, {2008}.

\bibitem{Ashhab2009}
S.~Ashhab, J.~R. Johansson, A.~M. Zagoskin, and Franco Nori.
\newblock {Single-artificial-atom lasing using a voltage-biased superconducting
  charge qubit}.
\newblock {\em NJP}, {11}:023030, {2009}.

\bibitem{Birnbaum2005}
K.~M. Birnbaum, A.~Boca, R.~Miller, A.~D. Boozer, T.~E. Northup, and H.~J.
  Kimble.
\newblock Photon blockade in an optical cavity with one trapped atom.
\newblock {\em Nature}, 436:87--90, 2005.

\bibitem{Moon2005}
K~Moon and SM~Girvin.
\newblock {Theory of microwave parametric down-conversion and squeezing using
  circuit QED}.
\newblock {\em Phys. Rev. Lett.}, {95}:140504, {2005}.

\bibitem{Marquardt2007}
Florian Marquardt.
\newblock Efficient on-chip source of microwave photon pairs in superconducting
  circuit qed.
\newblock {\em Phys. Rev. B}, 76:205416, 2007.

\bibitem{Castellanos-Beltran2008}
M.~A. Castellanos-Beltran, K.~D. Irwin, G.~C. Hilton, L.~R. Vale, and K.~W.
  Lehnert.
\newblock Amplification and squeezing of quantum noise with a tunable josephson
  metamaterial.
\newblock {\em Nature Physics}, 4:929--931, 2008.

\bibitem{Bergeal2008}
N.~Bergeal, R.~Vijay, V.~E. Manucharyan, I.~Siddiqi, R.~J. Schoelkopf, S.~M.
  Girvin, and M.~H. Devoret.
\newblock Analog information processing at the quantum limit with a josephson
  ring modulator.
\newblock {\em arXiv.org:0805.3452}, 2008.

\bibitem{Gottesman2001}
Daniel Gottesman, Alexei Kitaev, and John Preskill.
\newblock Encoding a qubit in an oscillator.
\newblock {\em Phys. Rev. A}, 64:012310, 2001.

\bibitem{Braunstein2005}
Samuel~L. Braunstein and Peter van Loock.
\newblock Quantum information with continuous variables.
\newblock {\em Rev. Mod. Phys.}, 77:513--577, 2005.

\bibitem{Martinis2005}
JM~Martinis, KB~Cooper, R~McDermott, M~Steffen, M~Ansmann, KD~Osborn, K~Cicak,
  S~Oh, DP~Pappas, RW~Simmonds, and CC~Yu.
\newblock {Decoherence in Josephson qubits from dielectric loss}.
\newblock {\em Phys. Rev. Lett.}, {95}:210503, {2005}.

\bibitem{Averin1985}
D.V. Averin, A.B. Zorin, and K.K. Likharev.
\newblock Bloch oscillations in small josephson junctions.
\newblock {\em Sov. Phys. JETP}, 61:407, 1985.

\bibitem{Buttiker1987}
M.~B\"uttiker.
\newblock Zero-current persistent potential drop across small-capacitance
  josephson junctions.
\newblock {\em Phys. Rev. B}, 36:3548--3555, 1987.

\bibitem{Lafarge1993}
P.~Lafarge, P.~Joyez, D.~Esteve, C.~Urbina, and M.~H. Devoret.
\newblock Two-electron quantization of the charge on a superconductor.
\newblock {\em Nature}, 365:422--424, 1993.

\bibitem{BouchiatCPBPhysScr1998}
V.~Bouchiat, D.~Vion, P.~Joyez, D.~Esteve, and M.~H. Devoret.
\newblock {\em Phys. Scr.}, T76:165, 1998.

\bibitem{Nakamura1999}
Y.~Nakamura, Yu.~A. Pashkin, and J.~S. Tsai.
\newblock Coherent control of macroscopic quantum states in a
  single-cooper-pair box.
\newblock {\em Nature}, 398:786--788, 1999.

\bibitem{Vion2002}
D.~Vion, A.~Aassime, A.~Cottet, P.~Joyez, H.~Pothier, C.~Urbina, D.~Esteve, and
  M.~H. Devoret.
\newblock {Manipulating the Quantum State of an Electrical Circuit}.
\newblock {\em Science}, 296:886--889, 2002.

\bibitem{Mooij1999}
J.~E. Mooij, T.~P. Orlando, L.~Levitov, Lin Tian, Caspar~H. van~der Wal, and
  Seth Lloyd.
\newblock {Josephson Persistent-Current Qubit}.
\newblock {\em Science}, 285:1036--1039, 1999.

\bibitem{Wal2000}
Caspar~H. van~der Wal, A.~C.~J. ter Haar, F.~K. Wilhelm, R.~N. Schouten, C.~J.
  P.~M. Harmans, T.~P. Orlando, Seth Lloyd, and J.~E. Mooij.
\newblock Quantum superposition of macroscopic persistent-current states.
\newblock {\em Science}, 290:773--777, 2000.

\bibitem{Chiorescu2003}
I.~Chiorescu, Y.~Nakamura, C.~J. P.~M. Harmans, and J.~E. Mooij.
\newblock {Coherent Quantum Dynamics of a Superconducting Flux Qubit}.
\newblock {\em Science}, 299:{1869--1871}, 2003.

\bibitem{Martinis2002}
JM~Martinis, S~Nam, J~Aumentado, and C~Urbina.
\newblock {Rabi oscillations in a large Josephson-junction qubit}.
\newblock {\em Phys. Rev. Lett.}, {89}:117901, {2002}.

\bibitem{Berkley2003}
A.~J. Berkley, H.~Xu, R.~C. Ramos, M.~A. Gubrud, F.~W. Strauch, P.~R. Johnson,
  J.~R. Anderson, A.~J. Dragt, C.~J. Lobb, and F.~C. Wellstood.
\newblock {Entangled Macroscopic Quantum States in Two Superconducting Qubits}.
\newblock {\em Science}, 300:1548--1550, 2003.

\bibitem{Aassime2001}
A.~Aassime, G.~Johansson, G.~Wendin, R.~J. Schoelkopf, and P.~Delsing.
\newblock Radio-frequency single-electron transistor as readout device for
  qubits: Charge sensitivity and backaction.
\newblock {\em Phys. Rev. Lett.}, 86:3376--3379, 2001.

\bibitem{Lehnert2003}
K.~W. Lehnert, K.~Bladh, L.~F. Spietz, D.~Gunnarsson, D.~I. Schuster,
  P.~Delsing, and R.~J. Schoelkopf.
\newblock Measurement of the excited-state lifetime of a microelectronic
  circuit.
\newblock {\em Phys. Rev. Lett.}, 90:027002, 2003.

\bibitem{Martinis2003}
JM~Martinis, S~Nam, J~Aumentado, KM~Lang, and C~Urbina.
\newblock {Decoherence of a superconducting qubit due to bias noise}.
\newblock {\em Phys. Rev. B}, {67}:094510, {2003}.

\bibitem{Widom1984}
A.~Widom, G.~Megaloudis, T.D. Clark, J.E. Mutton, R.J. Prance, and H.~Prance.
\newblock The josephson pendulum as a nonlinear capacitor.
\newblock {\em J. Low Temp. Phys.}, 57:651, 1984.

\bibitem{LiharevandZorin1985}
K.K. Likharev and A.B. Zorin.
\newblock Theory of the bloch-wave oscillations in small josephson junctions.
\newblock {\em J. Low Temp. Phys.}, 59:347, 1985.

\bibitem{Averin2003}
D.~V. Averin and C.~Bruder.
\newblock Variable electrostatic transformer: Controllable coupling of two
  charge qubits.
\newblock {\em Phys. Rev. Lett.}, 91:057003, 2003.

\bibitem{Duty2005}
T~Duty, G~Johansson, K~Bladh, D~Gunnarsson, C~Wilson, and P~Delsing.
\newblock {Observation of quantum capacitance in the cooper-pair transistor}.
\newblock {\em Phys. Rev. Lett.}, {95}:206807, {2005}.

\bibitem{Sillanpaa2005}
MA~Sillanp\"a\"a, T~Lehtinen, A~Paila, Y~Makhlin, L~Roschier, and PJ~Hakonen.
\newblock {Direct observation of Josephson capacitance}.
\newblock {\em Phys. Rev. Lett.}, {95}:206806, {2005}.

\bibitem{LifeAfterChargeNoise2009}
A.A. Houck, Jens Koch, M.H. Devoret, S.M. Girvin, and R.J. Schoelkopf.
\newblock Life after charge noise: recent results with transmon qubits.
\newblock {\em Quantum Information Processing}, 8:105, 2009.

\bibitem{Steffen2003}
M~Steffen, JM~Martinis, and IL~Chuang.
\newblock {Accurate control of Josephson phase qubits}.
\newblock {\em Phys. Rev. B}, {68}:224518, {2003}.

\bibitem{Lucero2008}
Erik Lucero, M.~Hofheinz, M.~Ansmann, Radoslaw~C. Bialczak, N.~Katz, Matthew
  Neeley, A.~D. O'Connell, H.~Wang, A.~N. Cleland, and John~M. Martinis.
\newblock {High-fidelity gates in a single Josephson qubit}.
\newblock {\em Phys. Rev. Lett.}, {100}:247001, {2008}.

\bibitem{Lutchyn2005}
R~Lutchyn, L~Glazman, and A~Larkin.
\newblock {Quasiparticle decay rate of Josephson charge qubit oscillations}.
\newblock {\em Phys. Rev. B}, {72}:014517, {2005}.

\bibitem{Lutchyn2006}
R.~M. Lutchyn, L.~I. Glazman, and A.~I. Larkin.
\newblock Kinetics of the superconducting charge qubit in the presence of a
  quasiparticle.
\newblock {\em Phys. Rev. B}, 74:064515, 2006.

\bibitem{MartinisQP2009}
John~M. Martinis, M.~Ansmann, and J.~Aumentado.
\newblock Energy decay in josephson qubits from non-equilibrium quasiparticles.
\newblock {\em arXiv:0904.2171}, 2009.

\bibitem{MartinisQP2009a}
John~M. Martinis.
\newblock Quasiparticle tunneling: P(e) theory.
\newblock {\em arXiv:0904.2035}, 2009.

\bibitem{Simmonds2004}
RW~Simmonds, KM~Lang, DA~Hite, S~Nam, DP~Pappas, and JM~Martinis.
\newblock {Decoherence in Josephson phase qubits from Junction resonators}.
\newblock {\em Phys. Rev. Lett.}, {93}:077003, {2004}.

\bibitem{Cooper2004}
KB~Cooper, M~Steffen, R~McDermott, RW~Simmonds, S~Oh, DA~Hite, DP~Pappas, and
  JM~Martinis.
\newblock {Observation of quantum oscillations between a Josephson phase qubit
  and a microscopic resonator using fast readout}.
\newblock {\em Phys. Rev. Lett.}, {93}:180401, {2004}.

\bibitem{Faoro2006}
Lara Faoro and Lev~B. Ioffe.
\newblock Quantum two level systems and kondo-like traps as possible sources of
  decoherence in superconducting qubits.
\newblock {\em Phys. Rev. Lett.}, 96:047001, 2006.

\bibitem{Faoro2007}
Lara Faoro and Lev~B. Ioffe.
\newblock Microscopic origin of critical current fluctuations in large, small,
  and ultra-small area josephson junctions.
\newblock {\em Phys. Rev. B}, 75:132505, 2007.

\bibitem{Neeley2008}
Matthew Neeley, M.~Ansmann, Radoslaw~C. Bialczak, M.~Hofheinz, N.~Katz, Erik
  Lucero, A.~O'Connell, H.~Wang, A.~N. Cleland, and John~M. Martinis.
\newblock {Process tomography of quantum memory in a Josephson-phase qubit
  coupled to a two-level state}.
\newblock {\em Nature Physics}, {4}:{523--526}, {2008}.

\bibitem{Constantin2009}
Magdalena Constantin, Clare~C. Yu, and John~M. Martinis.
\newblock Saturation of two-level systems and charge noise in josephson
  junction qubits.
\newblock {\em Phys. Rev. B}, 79:094520, 2009.

\bibitem{fluxonium}
Vladimir~E. Manucharyan, Jens Koch, Leonid Glazman, and Michel Devoret.
\newblock Defying the fine structure constant: single cooper pair circuit free
  of charge offsets.
\newblock {\em Science}, (in press):[arXiv:0906.0831v1], (2008).

\bibitem{Koch2005}
R.~H. Koch, J.~R. Rozen, G.~A. Keefe, F.~M. Milliken, C.~C. Tsuei, J.~R.
  Kirtley, and D.~P. DiVincenzo.
\newblock Low-bandwidth control scheme for an oscillator-stabilized josephson
  qubit.
\newblock {\em Phys. Rev. B}, 72:092512, 2005.

\bibitem{Koch2006}
R.~H. Koch, G.~A. Keefe, F.~P. Milliken, J.~R. Rozen, C.~C. Tsuei, J.~R.
  Kirtley, and D.~P. DiVincenzo.
\newblock Experimental demonstration of an oscillator stabilized josephson flux
  qubit.
\newblock {\em Phys. Rev. Lett.}, 96:127001, 2006.

\bibitem{Andre2006}
A.~Andre, D.~Demille, J.~M. Doyle, M.~D. Lukin, S.~E. Maxwell, P.~Rabl, R.~J.
  Schoelkopf, and P.~Zoller.
\newblock {A coherent all-electrical interface between polar molecules and
  mesoscopic superconducting resonators}.
\newblock {\em Nature Physics}, {2}:{636--642}, {2006}.

\bibitem{Schustermolionstrap2009}
D.~I. Schuster, Lev~S. Bishop, I.~L. Chuang, D.~DeMille, and R.~J. Schoelkopf.
\newblock Cavity qed in a molecular ion trap.
\newblock {\em arXiv:0903.355}, 2009.

\bibitem{Gambetta2007}
Jay Gambetta, W.~A. Braff, A.~Wallraff, S.~M. Girvin, and R.~J. Schoelkopf.
\newblock {Protocols for optimal readout of qubits using a continuous quantum
  nondemolition measurement}.
\newblock {\em Phys. Rev. A}, {76}:012325, {2007}.

\bibitem{Mallet2009}
F.~Mallet, F.~Ong, A.~Palacios-Laloy, F.~Nguyen, P.~Bertet, D.~Vion, and
  D.~Esteve.
\newblock Single-shot qubit readout for circuit quantum electrodynamics.
\newblock {\em Nature Physics}, (in press), 2009.

\bibitem{Lupascu2007}
A.~Lupascu, S.~Saito, T.~Picot, P.~C. De~Groot, C.~J. P.~M. Harmans, and J.~E.
  Mooij.
\newblock {Quantum non-demolition measurement of a superconducting two-level
  system}.
\newblock {\em Nature Physics}, {3}:{119--123}, {2007}.

\bibitem{Katz2006}
N~Katz, M~Ansmann, RC~Bialczak, E~Lucero, R~McDermott, M~Neeley, M~Steffen,
  EM~Weig, AN~Cleland, JM~Martinis, and AN~Korotkov.
\newblock {Coherent state evolution in a superconducting qubit from
  partial-collapse measurement}.
\newblock {\em Science}, {312}:{1498--1500}, {2006}.

\bibitem{Gambetta2008}
Jay Gambetta, Alexandre Blais, M.~Boissonneault, A.~A. Houck, D.~I. Schuster,
  and S.~M. Girvin.
\newblock Quantum trajectory approach to circuit qed: Quantum jumps and the
  zeno effect.
\newblock {\em Physical Review A}, 77:012112, 2008.

\bibitem{Steffen2006a}
Matthias Steffen, M.~Ansmann, Radoslaw~C. Bialczak, N.~Katz, Erik Lucero,
  R.~McDermott, Matthew Neeley, E.~M. Weig, A.~N. Cleland, and John~M.
  Martinis.
\newblock {Measurement of the entanglement of two superconducting qubits via
  state tomography}.
\newblock {\em Science}, {313}:{1423--1425}, {2006}.

\bibitem{UCSB_Bell2009}
Markus Ansmann, H.~Wang, Radoslaw~C. Bialczak, Max Hofheinz, Erik Lucero,
  M.~Neeley, A.~D. O'Connell, D.~Sank, M.~Weides, J.~Wenner, A.~N. Cleland, and
  John~M. Martinis.
\newblock Violation of bell's inequality in josephson phase qubits.
\newblock {\em Nature}, 461:504--506, 2009.

\bibitem{Bishop2009a}
Lev~S Bishop, L~Tornberg, D~Price, E~Ginossar, A~Nunnenkamp, A~A Houck, J~M
  Gambetta, Jens Koch, G~Johansson, S~M Girvin, and R~J Schoelkopf.
\newblock Proposal for generating and detecting multi-qubit ghz states in
  circuit qed.
\newblock {\em NJP}, 11:073040, 2009.

\bibitem{Filipp2009}
S.~Filipp, P.~Maurer, P.~J. Leek, M.~Baur, R.~Bianchetti, J.~M. Fink,
  M.~G\"{o}ppl, L.~Steffen, J.~M. Gambetta, A.~Blais, and A.~Wallraff.
\newblock Two-qubit state tomography using a joint dispersive readout.
\newblock {\em Physical Review Letters}, 102:200402, 2009.

\bibitem{EntanglementMetrology2009}
J.~M. Chow, L.~DiCarlo, J.~M. Gambetta, A.~Nunnenkamp, Lev~S. Bishop,
  L.~Frunzio, M.~H. Devoret, S.~M. Girvin, and R.~J. Schoelkopf.
\newblock Entanglement metrology using a joint readout of superconducting
  qubits.
\newblock (arXiv:0908.1955).

\bibitem{Neeley2009}
Matthew Neeley, Markus Ansmann, Radoslaw~C. Bialczak, Max Hofheinz, Erik
  Lucero, Aaron~D. O'Connell, Daniel Sank, Haohua Wang, James Wenner, Andrew~N.
  Cleland, Michael~R. Geller, and John~M. Martinis.
\newblock {Emulation of a Quantum Spin with a Superconducting Phase Qudit}.
\newblock {\em Science}, 325:722--725, 2009.

\bibitem{Steffen2006}
Matthias Steffen, M.~Ansmann, R.~McDermott, N.~Katz, Radoslaw~C. Bialczak, Erik
  Lucero, Matthew Neeley, E.~M. Weig, A.~N. Cleland, and John~M. Martinis.
\newblock {State tomography of capacitively shunted phase qubits with high
  fidelity}.
\newblock {\em Phys. Rev. Lett.}, {97}:050502, {2006}.

\bibitem{Rigetti2005}
C~Rigetti, A~Blais, and M~Devoret.
\newblock {Protocol for universal gates in optimally biased superconducting
  qubits}.
\newblock {\em Phys. Rev. Lett.}, {94}:240502, {2005}.

\bibitem{Bertet2006}
P~Bertet, CJPM Harmans, and JE~Mooij.
\newblock {Parametric coupling for superconducting qubits}.
\newblock {\em Phys. Rev. B}, {73}:064512, {2006}.

\bibitem{Gywat2006}
O~Gywat, F~Meier, D~Loss, and DD~Awschalom.
\newblock {Dynamics of coupled qubits interacting with an off-resonant cavity}.
\newblock {\em Phys. Rev. B}, {73}:125336, {2006}.

\bibitem{Blais2007}
Alexandre Blais, Jay Gambetta, A.~Wallraff, D.~I. Schuster, S.~M. Girvin, M.~H.
  Devoret, and R.~J. Schoelkopf.
\newblock {Quantum-information processing with circuit quantum
  electrodynamics}.
\newblock {\em Phys. Rev. A}, {75}:032329, {2007}.

\bibitem{Plantenberg2007}
J.~H. Plantenberg, P.~C. de~Groot, C.~J. P.~M. Harmans, and J.~E. Mooij.
\newblock Demonstration of controlled-not quantum gates on a pair of
  superconducting quantum bits.
\newblock {\em Nature}, 447:836--839, 2007.

\bibitem{Pashkin2003}
Yu.~A. Pashkin, T.~Yamamoto, O.~Astafiev, Y.~Nakamura, D.~V. Averin, and J.~S.
  Tsai.
\newblock Quantum oscillations in two coupled charge qubits.
\newblock {\em Nature}, 421:823--826, 2003.

\bibitem{Yamamoto2003}
T.~Yamamoto, Yu.~A. Pashkin, O.~Astafiev, Y.~Nakamura, and J.~S. Tsai.
\newblock Demonstration of conditional gate operation using superconducting
  charge qubits.
\newblock {\em Nature}, 425:941--944, 2003.

\bibitem{McDermott2005}
R~McDermott, RW~Simmonds, M~Steffen, KB~Cooper, K~Cicak, KD~Osborn, S~Oh,
  DP~Pappas, and JM~Martinis.
\newblock {Simultaneous state measurement of coupled Josephson phase qubits}.
\newblock {\em Science}, {307}:{1299--1302}, {2005}.

\bibitem{Majer2005}
J.~B. Majer, F.~G. Paauw, A.~C.~J. ter Haar, C.~J. P.~M. Harmans, and J.~E.
  Mooij.
\newblock Spectroscopy on two coupled superconducting flux qubits.
\newblock {\em Phys. Rev. Lett.}, 94:090501, 2005.

\bibitem{Strauch2003}
Frederick~W. Strauch, Philip~R. Johnson, Alex~J. Dragt, C.~J. Lobb, J.~R.
  Anderson, and F.~C. Wellstood.
\newblock Quantum logic gates for coupled superconducting phase qubits.
\newblock {\em Phys. Rev. Lett.}, 91:167005, 2003.

\bibitem{Ke-Hui2007}
Song Ke-Hui, Zhou Zheng-Wei, and Guo Guang-Can.
\newblock {Implementation of a controlled-phase gate and Deutsch-Jozsa
  algorithm with superconducting charge qubits in a cavity}.
\newblock {\em Comm. Theor. Phys.}, {47}:{821--825}, {2007}.

\bibitem{Wallquist2006}
M.~Wallquist, V.~S. Shumeiko, and G.~Wendin.
\newblock Selective coupling of superconducting charge qubits mediated by a
  tunable stripline cavity.
\newblock {\em Phys. Rev. B}, 74:224506, 2006.

\bibitem{Palacios-Laloy2008}
A.~Palacios-Laloy, F.~Nguyen, F.~Mallet, P.~Bertet, D.~Vion, and D.~Esteve.
\newblock Tunable resonators for quantum circuits.
\newblock {\em J. Low Temp. Phys.}, 151:1034--1042, 2008.

\bibitem{Hime2006}
T.~Hime, P.~A. Reichardt, B.~L.~T. Plourde, T.~L. Robertson, C.-E. Wu, A.~V.
  Ustinov, and John Clarke.
\newblock {Solid-State Qubits with Current-Controlled Coupling}.
\newblock {\em Science}, 314:1427--1429, 2006.

\bibitem{Niskanen2007}
A.~O. Niskanen, K.~Harrabi, F.~Yoshihara, Y.~Nakamura, S.~Lloyd, and J.~S.
  Tsai.
\newblock {Quantum Coherent Tunable Coupling of Superconducting Qubits}.
\newblock {\em Science}, 316:723--726, 2007.

\bibitem{SMGRJSunpublished}
S.M. Girvin and R.J. Schoelkopf.
\newblock (unpublished).

\bibitem{DobsicekPhaseEstimation2007}
M.~Dobsicek, G.~Johansson, V.S. Shumeiko, and G.~Wendin:.
\newblock Arbitrary accuracy iterative phase estimation algorithm as a two
  qubit benchmark.
\newblock {\em Phys. Rev. A}, 76:030306(R), 2007.

\bibitem{Wei2006}
L.~F. Wei, {Yu-xi} Liu, and Franco Nori.
\newblock Generation and control of greenberger-horne-zeilinger entanglement in
  superconducting circuits.
\newblock {\em Phys. Rev. Lett.}, 96:246803, 2006.

\bibitem{Tsomokos2008}
Dimitris~I. Tsomokos, Sahel Ashhab, and Franco Nori.
\newblock {Fully connected network of superconducting qubits in a cavity}.
\newblock {\em NJP}, {10}:113020, {2008}.

\bibitem{Galiautdinov2008}
Andrei Galiautdinov and John~M. Martinis.
\newblock Maximally entangling tripartite protocols for josephson phase qubits.
\newblock {\em Phys. Rev. A}, 78:010305(R), 2008.

\bibitem{MikeandIke}
I.~L. Chuang and M.~A. Nielsen.
\newblock {\em Quantum Information and Quantum Computation}.
\newblock Cambridge University Press, 2000.

\bibitem{TornbergPhaseFlip2008}
L.~Tornberg, M.~Wallquist, G.~Johansson, V.S. Shumeiko, and G.~Wendin.
\newblock Implementation of the three-qubit phase-flip error correction code
  with superconducting qubits.
\newblock {\em Phys. Rev. B 77}, 77:214528, 2008.

\bibitem{Kitaev2003}
A.Yu. Kitaev.
\newblock Fault-tolerant quantum computation by anyons.
\newblock {\em Ann. Phys.}, 303:2--30, 2003.

\bibitem{Ioffe2002}
L.~B. Ioffe and M.~V. Feigelman.
\newblock Possible realization of an ideal quantum computer in josephson
  junction array.
\newblock {\em Phys. Rev. B}, 66:224503, 2002.

\bibitem{Ioffe2002a}
L.~B. Ioffe, M.~V. Feigel'man, A.~Ioselevich, D.~Ivanov, M.~Troyer, and
  G.~Blatter.
\newblock Topologically protected quantum bits using josephson junction arrays.
\newblock {\em Nature}, 415:503--506, 2002.

\bibitem{Doucot2003}
B.~Doucot, M.~V. Feigelman, and L.~B. Ioffe.
\newblock Topological order in the insulating josephson junction arrays.
\newblock {\em Phys. Rev. Lett.}, 90:107003, 2003.

\bibitem{Doucot2005}
B.~Doucot, M.~V. Feigelman, L.~B. Ioffe, and A.~S. Ioselevich.
\newblock Protected qubits and chern-simons theories in josephson junction
  arrays.
\newblock {\em Phys. Rev. B}, 71:024505, 2005.

\bibitem{Gladchenko2009}
Sergey Gladchenko, David Olaya, Eva Dupont-Ferrier, Benoit Doucot, Lev~B.
  Ioffe, and Michael~E. Gershenson.
\newblock Superconducting nanocircuits for topologically protected qubits.
\newblock {\em Nature Physics}, 5:48--53, 2009.

\bibitem{Ogden2008}
C.~D. Ogden, E.~K. Irish, and M.~S. Kim.
\newblock {Dynamics in a coupled-cavity array}.
\newblock {\em Phys. Rev. A}, {78}:063805, {2008}.

\bibitem{Helmer2009}
F.~Helmer, M.~Mariantoni, A.~G. Fowler, J.~von Delft, E.~Solano, and
  F.~Marquardt.
\newblock {Cavity grid for scalable quantum computation with superconducting
  circuits}.
\newblock {\em {EPL}}, {85}:50007, {2009}.

\bibitem{Fisher1989}
Matthew P.~A. Fisher, Peter~B. Weichman, G.~Grinstein, and Daniel~S. Fisher.
\newblock Boson localization and the superfluid-insulator transition.
\newblock {\em Phys. Rev. B}, 40:546--570, 1989.

\bibitem{Greentree2006}
Andrew~D. Greentree, Charles Tahan, Jared~H. Cole, and Lloyd C.~L. Hollenberg.
\newblock {Quantum phase transitions of light}.
\newblock {\em {Nature Physics}}, {2}:{856--861}, {2006}.

\bibitem{Illuminati2006}
Fabrizio Illuminati.
\newblock Quantum optics: Light does matter.
\newblock {\em Nature Physics}, 2:803--804, 2006.

\bibitem{Hartmann2007}
Michael~J. Hartmann and Martin~B. Plenio.
\newblock {Strong photon nonlinearities and photonic mott insulators}.
\newblock {\em Phys. Rev. Lett.}, {99}:103601, {2007}.

\bibitem{Jarrett2007}
T.~C. Jarrett, A.~Olaya-Castro, and N.~F. Johnson.
\newblock {Optical signatures of quantum phase transitions in a light-matter
  system}.
\newblock {\em {EPL}}, {77}:34001, {2007}.

\bibitem{Rossini2007}
Davide Rossini and Rosario Fazio.
\newblock Mott-insulating and glassy phases of polaritons in 1d arrays of
  coupled cavities.
\newblock {\em Phys. Rev. Lett.}, 99:186401, 2007.

\bibitem{Hartmann2008}
M~J Hartmann, F~G S~L Brandao, and M~B Plenio.
\newblock A polaritonic two-component bose-hubbard model.
\newblock {\em NJP}, 10:033011 (12pp), 2008.

\bibitem{Makin2008}
M.~I. Makin, Jared~H. Cole, Charles Tahan, Lloyd C.~L. Hollenberg, and
  Andrew~D. Greentree.
\newblock Quantum phase transitions in photonic cavities with two-level
  systems.
\newblock {\em Phys. Rev. A}, {77}:053819, {2008}.

\bibitem{Aichhorn2008}
Markus Aichhorn, Martin Hohenadler, Charles Tahan, and Peter~B. Littlewood.
\newblock Quantum fluctuations, temperature, and detuning effects in
  solid-light systems.
\newblock {\em Phys. Rev. Lett.}, 100:216401, 2008.

\bibitem{Cho2008}
Jaeyoon Cho, Dimitris~G. Angelakis, and Sougato Bose.
\newblock Simulation of high-spin heisenberg models in coupled cavities.
\newblock {\em Phys. Rev. A}, 78:062338, 2008.

\bibitem{Cho2008a}
Jaeyoon Cho, Dimitris~G. Angelakis, and Sougato Bose.
\newblock Fractional quantum hall state in coupled cavities.
\newblock {\em Phys. Rev. Lett.}, 101:246809, 2008.

\bibitem{Zhao2008}
J.~Zhao, A.~W. Sandvik, and K.~Ueda.
\newblock Insulator to superfluid transition in coupled photonic cavities in
  two dimensions.
\newblock {\em arXiv:0806.3603}, 2008.

\bibitem{Na2008}
Neil Na, Shoko Utsunomiya, Lin Tian, and Yoshihisa Yamamoto.
\newblock Strongly correlated polaritons in a two-dimensional array of photonic
  crystal microcavities.
\newblock {\em Phys. Rev. A}, 77:031803, 2008.

\bibitem{Lei2008}
Soi-Chan Lei and Ray-Kuang Lee.
\newblock Quantum phase transitions of light in the dicke-bose-hubbard model.
\newblock {\em Phys. Rev. A}, 77:033827, 2008.

\bibitem{Ji2009}
An-Chun Ji, Qing Sun, X.~C. Xie, and W.~M. Liu.
\newblock Josephson effect for photons in two weakly linked microcavities.
\newblock {\em Phys. Rev. Lett.}, 102:023602, 2009.

\bibitem{LaserPhotonicsReview2009}
J.~Hartmann, F.G.S.L. Brandao, and M.B. Plenio.
\newblock Quantum many-body phenomena in coupled cavity arrays.
\newblock {\em Laser \& Photonics Review}, 2:527--556, 2009.

\bibitem{Grochol2009}
Michal Grochol.
\newblock Quantum phase transitions in an array of coupled nanocavity quantum
  dots.
\newblock {\em Phys. Rev. B}, 79:205306, 2009.

\bibitem{Dalidovich2009}
Denis Dalidovich and Malcolm~P. Kennett.
\newblock Bose-hubbard model in the presence of ohmic dissipation.
\newblock {\em Phys. Rev. A}, 79:053611, 2009.

\bibitem{Carusotto2009}
I.~Carusotto, D.~Gerace, H.~E. Tureci, S.~De Liberato, C.~Ciuti, and
  A.~Imamoglu.
\newblock Fermionized photons in an array of driven dissipative nonlinear
  cavities.
\newblock {\em Phys. Rev. Lett.}, 103:033601, 2009.

\bibitem{Schmidt2009}
S.~Schmidt and G.~Blatter.
\newblock Strong coupling theory for the jaynes-cummings-hubbard model.
\newblock {\em arXiv:0905.3344}, 2009.

\bibitem{Koch2009}
Jens Koch and Karyn~Le Hur.
\newblock Superfluid--mott insulator transition of light in the jaynes-cummings
  lattice.
\newblock {\em Phys. Rev. A}, 80:023811, 2009.

\end{thebibliography}
%
\end{document}